\documentclass[aps,prd,twocolumn,floatfix,altaffilletter,superscriptaddress,preprintnumbers,
               tightenlines,showpacs,showkeys,nofootinbib,notoccite]{revtex4-1}
\usepackage[utf8]{inputenc}
\usepackage[colorlinks=true,citecolor=blue,linkcolor=blue]{hyperref}
\usepackage[normalem]{ulem}
\usepackage{amsmath,amssymb, mathrsfs,esvect}
\usepackage{epsfig}
\usepackage{graphicx}               
\usepackage{url}
\usepackage{color}
\usepackage{slashed}
\usepackage{multirow}
\usepackage{placeins}
\usepackage[dvipsnames]{xcolor}
\usepackage{epstopdf}
\usepackage{soul}
\usepackage{tikz}
\usepackage[capitalise, english]{cleveref}
\usepackage{siunitx}
\usepackage{xspace}
\usepackage{booktabs}
\usetikzlibrary{trees}
\usetikzlibrary{decorations.pathmorphing}
\usetikzlibrary{decorations.markings}

\newcommand\myshade{80}
\colorlet{mylinkcolor}{Blue}
\colorlet{mycitecolor}{Red}
\colorlet{myurlcolor}{violet}

\hypersetup{
  linkcolor  = mylinkcolor!\myshade!black,
  citecolor  = mycitecolor!\myshade!black,
  urlcolor   = myurlcolor!\myshade!black,
  colorlinks = true
}

\definecolor{jblue}{RGB}{20,50,100}
\definecolor{npurple}{RGB} {153, 51, 204}
\definecolor{wred}{RGB}{217,0,56}
\definecolor{white}{RGB}{255,255,255}

\definecolor{korange}{RGB}{235, 80,  43}
\definecolor{korange2}{RGB}{245, 100,  63}
\definecolor{kyelloworange}{RGB}{255, 210,  110}
\definecolor{kyelloworange2}{RGB}{240, 170,  90}
\definecolor{kred}{RGB}{204,  102, 153}
\definecolor{kpurple}{RGB}{153,  61, 190}
\definecolor{kpurplelight}{RGB}{213,  161, 230}

\allowdisplaybreaks

\providecommand{\abs}[1]{\lvert#1\rvert}

\def\thesection{\Roman{section}}

\usepackage{tikz,xcolor,hyperref}

\definecolor{lime}{HTML}{A6CE39}
\DeclareRobustCommand{\orcidicon}{\hspace{-1mm}
	\begin{tikzpicture}
	\draw[lime, fill=lime] (0,0) 
	circle [radius=0.16] 
	node[white] {{\fontfamily{qag}\selectfont \tiny \,ID}};
	\draw[white, fill=white] (-0.0525,0.095) 
	circle [radius=0.007];
	\end{tikzpicture}
	\hspace{-3mm}
}

\foreach \x in {A, ..., Z}{\expandafter\xdef\csname orcid\x\endcsname{\noexpand\href{https://orcid.org/\csname orcidauthor\x\endcsname}
		{\noexpand\orcidicon}}
}


\begin{document}

\title{Near future MeV telescopes can discover \\ asteroid-mass primordial black hole dark matter}

\author{Anupam Ray\orcidA{}}
\email{anupam.ray@theory.tifr.res.in}
\affiliation{Tata Institute of Fundamental Research, Homi Bhabha
	Road, Mumbai 400005, India}

\author{Ranjan Laha\orcidB{}} 
\email{ranjanlaha@iisc.ac.in}
\affiliation{Centre for High Energy Physics, Indian Institute of Science, C.\,V.\,Raman Avenue, Bengaluru 560012, India}

\author{Julian B.\,Mu$\tilde{\rm{n}}$oz\orcidC{}}
\email{julianmunoz@cfa.harvard.edu}
\affiliation{Harvard-Smithsonian Center for Astrophysics, 60 Garden St., Cambridge, Massachusetts 02138, USA}

\author{Regina Caputo\orcidD{}} 
\email{regina.caputo@nasa.gov}
\affiliation{NASA Goddard Space Flight Center, Greenbelt, Maryland 20771, USA}

\date{\today}


\begin{abstract}
Primordial black holes (PBHs), formed out of large overdensities in the early Universe, are a viable dark matter (DM) candidate over a broad range of masses.  
Ultra-light, asteroid-mass PBHs with masses around $10^{17}$\,g
are particularly interesting as current observations allow them to constitute the entire DM density. 
PBHs in this mass range emit $\sim$\,MeV photons via Hawking radiation which can directly be detected by the gamma ray telescopes, such as the upcoming AMEGO. 
In this work we forecast how well an instrument with the sensitivity of AMEGO will be able to detect, or rule out, PBHs as a DM candidate, by searching for their evaporating signature when marginalizing over the Galactic and extra-Galactic gamma-ray backgrounds.
We find that an instrument with the sensitivity of AMEGO could exclude non-rotating PBHs as the only DM component for masses up to $7 \times 10^{17}$ g  at 95\% confidence level (C.L.) for a monochromatic mass distribution, improving upon current bounds by nearly an order of magnitude.
The forecasted constraints are more stringent for PBHs that have rotation, or which follow extended mass distributions.
\end{abstract}

\maketitle
\preprint{TIFR/TH/21-1}

\section{Introduction} 
Unequivocal evidence of a non-baryonic form of matter, known as dark matter (DM), as a dominant component of the Universe has been confirmed by numerous astrophysical and cosmological observations~\cite{Aghanim:2018eyx,Pardo:2020epc,Strigari:2013iaa}. Experimental searches for the elusive DM have thus far shown no firmly preferred model~\cite{Slatyer:2017sev,Lin:2019uvt,Boveia:2018yeb}. Primordial black holes (PBHs), possibly formed via gravitational collapse of large overdensities  in the early universe or via other exotic mechanisms, are one of the earliest proposed and well-motivated DM candidates~\cite{1966AZh....43..758Z,Hawking:1971ei,Carr:1974nx,Chapline:1975ojl}.  
PBHs have a wide range of masses and can constitute a large fraction or even the entirety of the DM density~\cite{Katz:2018zrn,Montero-Camacho:2019jte,Smyth:2019whb,Dasgupta:2019cae,Carr:2020gox,Green:2020jor,Carr:2020xqk}. 
The idea of PBH DM has recently received renewed attention with the first detection of a BH merger by the LIGO-Virgo collaboration~\cite{Abbott:2016blz}, 
argued to have a PBH rather than astrophysical origin~\cite{Bird:2016dcv,Clesse:2016vqa,Sasaki:2016jop}. Several techniques have been implemented to probe the DM fraction of PBHs over a wide mass range.  These have resulted in 
a multitude of observational constraints~\cite{Arbey:2019vqx, Clark:2016nst, Wang:2016ana,Poulter:2019ooo, Boudaud:2018hqb, DeRocco:2019fjq, Laha:2019ssq, Ballesteros:2019exr,Dasgupta:2019cae, Smyth:2019whb, Laha:2020ivk, Allsman:2000kg, Tisserand:2006zx, Niikura:2019kqi, Oguri:2017ock, Zumalacarregui:2017qqd, Authors:2019qbw, Kavanagh:2018ggo, Brandt:2016aco, Koushiappas:2017chw,Monroy-Rodriguez:2014ula, Serpico:2020ehh, Hektor:2018qqw, Manshanden:2018tze, Hektor:2018rul,Raidal:2017mfl, Sammons:2020kyk, Lu:2020bmd,Nitz:2020bdb,Nitz:2021mzz,Laha:2020vhg,2021MNRAS.504.5475K,Chan:2020zry,Dolgov:2020xzo,Wong:2020yig,Hutsi:2020sol,Coogan:2020tuf,Halder:2021jiv}, along with several future projections~\cite{Munoz:2016tmg, Laha:2018zav, Katz:2019qug, Jung:2017flg, Kuhnel:2018mlr, Cai:2018dig, Jung:2019fcs, Katz:2018zrn,Bai:2018bej,Ballesteros:2019exr,Wang:2019kzb,Dror:2019twh,Guo:2017njn,Katz:2019qug,Dutta:2020lqc,Kusenko:2020pcg,Sugiyama:2020roc,Bhaumik:2020dor} along a broad range of PBH masses.

Due to their Hawking emission, extremely light PBHs would have evaporated by today, setting a lower limit on the mass  of $\sim$ $5 \times 10^{14}\,{\rm g}$ for non-rotating PBHs (or $\sim$ $7 \times 10^{14}\,{\rm g}$ if maximally rotating)~\cite{Page:1976df,Page:1976ki,MacGibbon:2007yq}.
PBHs heavier than that still evaporate, and  act as decaying DM.
Ultra-light PBHs with masses in between $5 \times 10^{14}\,{\rm g} -  2 \times 10^{17}\,{\rm g}$, are typically probed via searches of their Hawking radiation. 
Non-observations of such Hawking-produced photons~\cite{Arbey:2019vqx,Laha:2020ivk,Ballesteros:2019exr,Coogan:2020tuf}, neutrinos~\cite{Dasgupta:2019cae}, and electrons/\,positrons~\cite{Boudaud:2018hqb,1980AA....81..263O, okeke1980primary, 1991ApJ...371..447M, Bambi:2008kx,DeRocco:2019fjq,Laha:2019ssq,Dasgupta:2019cae} provide the leading constraints on ultra-light PBHs.  Additional constraints in this mass range are also obtained via precise observations of the cosmic microwave background and Big Bang Nucleosynthesis\,\cite{Stocker:2018avm,Acharya:2020jbv,Clark:2016nst,Keith:2020jww}.  PBHs in the mass range of $\sim$ $2 \times 10^{17}\, {\rm g} - 10^{23}\, {\rm g}$, often known as the asteroid-mass range\footnote{\href{https://nssdc.gsfc.nasa.gov/planetary/factsheet/asteroidfact.html}{https://nssdc.gsfc.nasa.gov/planetary/factsheet/asteroidfact.html}}, are currently allowed to compose the entirety of the DM~\cite{Katz:2018zrn,Montero-Camacho:2019jte,Smyth:2019whb}.  
Unlike solar-mass BHs, these ultra-light BHs cannot be produced by any known astrophysical processes (even with the continued accumulation of asymmetric DM particles in compact objects~\cite{Kouvaris:2018wnh,Dasgupta:2020mqg}), and thus would be a smoking gun of new physics, be it during the early Universe or in a complex dark sector~\cite{Shandera:2018xkn}.

In this work, we propose a technique to decisively probe a part of the parameter space for PBH DM in the asteroid-mass range.  
We show that observation of the Galactic Center by future MeV telescopes, such as an instrument with the sensitivity of  AMEGO~\cite{2019BAAS...51g.245M}, can probe the DM fraction of asteroid-mass PBHs.  AMEGO can exclude non-rotating (maximally rotating) PBHs as the sole component of DM upto $\sim$ $7 \times 10^{17}$ g ($\sim$ $4 \times 10^{18}$ g), at 95\% C.L., assuming no signal is present in the data and a monochromatic mass function of PBHs.  Assuming that PBHs follow an extended mass distribution (log-normal distribution with width $\sigma =0.5$), AMEGO can probe further into an entirely unexplored mass window, improving our current constraints by nearly an order of magnitude and pushing us closer to probe the entire asteroid-mass PBH window. Ref.~\cite{Coogan:2020tuf}, which appeared as our paper was near completion, performs a similar study for non-rotating PBHs with a monochromatic mass distribution. Our work differs from Ref.~\cite{Coogan:2020tuf} in several key aspects (e.g., the inclusion of the extra-Galactic astrophysical background, and a different region of interest in the Galactic Center), chief among them is our usage of Fisher analysis to derive the projected exclusion limits on the DM fraction of ultra-light PBHs including marginalization over the astrophysical parameters.

\section{Particle emission from evaporating black holes}

BHs evaporate via Hawking radiation~\cite{Hawking:1971ei}. An uncharged and rotating BH of mass $M_{\rm{BH}}$ and angular momentum $J_{\rm{BH}}$ radiates at a temperature~\cite{Page:1976df,Page:1976ki,MacGibbon:2007yq,MacGibbon:1990zk,MacGibbon:1991tj}
\begin{equation}
T_{\rm{BH}} = \frac{1}{4 \pi G_{N} M_{\rm{BH}}}  \frac{\sqrt{1-a_*^2}}{1+\sqrt{1-a_*^2}}\,,
\label{eq:PBH temperature}
\end{equation}
where $G_N$ denotes  the gravitational constant and $a_* = J_{\rm{BH}}/(G_N M^2_{\rm{BH}})$ is the dimensionless spin parameter. For a given BH mass, the temperature can vary by orders of magnitude as it approaches its maximal spin, $a_* \to 1$, where the BH stops evaporating.

The number of emitted particles from an evaporating BH of mass $M_{\rm{BH}}$ and dimensionless spin parameter $a_*$, in the energy interval $E$ and $E + dE$ and in a time interval $dt$ follows a blackbody-like distribution~\cite{Page:1976df,Page:1976ki,Hawking:1971ei,MacGibbon:2007yq,MacGibbon:1990zk,MacGibbon:1991tj}
\begin{equation}
\frac{d^2N}{dEdt} = \frac{1}{2\pi} \frac{\Gamma_s(E,M_{\rm{BH}},a_*,\mu)}{{\exp}\left[{E'}/{T_{\rm{BH}}} \right]-(-1)^{2s}} \,,
\label{eq:Differential energy distribution}
\end{equation} 
where  $E'$ denotes the effective energy of the emitted particles including the rotational velocity of the BH. $\Gamma_s$ denotes the graybody factor which accounts for the departure from an ideal blackbody emission. It depends on the spin of the emitted particle $s$, rest mass of the emitted particle $\mu$, and the BH mass and spin. In the high energy limit, $G_N M_{\rm{BH}} E \gg 1$, the graybody factor becomes independent of the spin of the emitted particle species and reaches its geometric saturation value, i.e.\, $\Gamma_s = 27 G^2_N M^2_{\rm{BH}} E^2$.  In the opposite limit, $G_N M_{\rm{BH}} E \ll 1$, it strongly depends on the spin of the emitted particle species~\cite{MacGibbon:1990zk,Page:1976df}.

For this work, we use the publicly available {\tt BlackHawk}~\cite{Arbey:2019mbc} package to generate the emitted particle spectrum from evaporating BHs. We have verified this numerically obtained emission rate against
semi-analytical formulae from Ref.~\cite{Page:1976df, Page:1976ki,MacGibbon:1990zk}. 

\section{Methods \& Results}

Hereafter we focus on primordial black holes. 
Ultra-light PBHs emit significant number of photons of energy comparable to their temperature. 
More precisely, photon emission peaks at an energy $E \sim 5.77\,T_{\rm{PBH}}$~\cite{MacGibbon:1990zk,MacGibbon:2007yq} for an evaporating PBH with temperature $T_{\rm{PBH}}$. 
The emission of photons is exponentially suppressed for energies exceeding $T_{\rm{PBH}}$ ($E \gg T_{\rm{PBH}}$), and falls off as a power law in the opposite limit ($E \ll T_{\rm{PBH}}$). 
 
For a monochromatic mass distribution of PBHs, the Galactic contribution to the differential flux from PBH evaporation is
  	\begin{equation}
 \frac{d\phi_{\rm gal}} {dE}\Bigr\rvert_{\rm{mono}}=  \frac{f_{\rm PBH}}{4\pi M_{\rm{PBH}}}   \frac{d^2N}{dE dt} \int_{0}^{s_{\rm{max}}} \rho\,[r(s,l,b)]\,ds\,d \Omega\,,
 \label{eq: Galactic photon contribution}
 \end{equation}
 where $f_{\rm PBH}$ denotes the DM fraction of PBHs. DM profile of the Milky Way (MW) is denoted by $\rho\,[r(s,l,b)]$, where $r$ is the Galacto-centric distance, $s$ is the distance from the observer, $l$ and $b$ denote the Galactic longitude and latitude respectively, and $d\Omega = \cos[b]\,dl\,db$ is the differential solid angle under consideration. The upper limit of the line of sight integral, $s_{\rm{max}}$, depends on the size of the MW DM halo, Galactic longitude, and Galactic latitude:
 \begin{equation}
 s_{\rm{max}} = r_{\odot}\,\cos[b]\,\cos[l]+\sqrt{ r^2_{\rm{max}}-r^2_{\odot} \left(1-\cos^2[b]\,\cos^2[l]\right) }\,,
 \label{eq: smax}
 \end{equation}
 where $r_{\rm{max}}$ denotes the maximum size of the MW halo, and $r_{\odot}$ is the Galacto-centric distance of the Sun.

The  extra-Galactic contribution to the differential flux for a monochromatic mass distribution of PBHs is
\begin{equation}
\frac{d\phi_{\rm eg}} {dE}\Bigr\rvert_{\rm{mono}} = \frac{\Delta \Omega}{4\pi} \frac{f_{\rm{PBH}}\,\rho_{\rm DM}}{M_{\rm{PBH}}}  \int_{z = 0}^\infty \frac{dz}{H(z)}\, \frac{d^2N}{dE dt}\Bigr\rvert_{E \to [1+z]E}\,,
\label{eq:extragalactic contribution}
\end{equation}
where $\Delta \Omega$ denotes the total solid angle under consideration, $\rho_{\rm{DM}}$ is the average DM density of the Universe at the present epoch, and $H(z) = H_0\,\sqrt{\Omega_{\Lambda}+ \Omega_{m}(1+z)^3+\Omega_{r}(1+z)^4}$ is the Hubble expansion rate at redshift $z$.  The Hubble expansion rate at the present epoch is $H_0$; $\Omega_{\Lambda},\Omega_{m}$, and $\Omega_{r}$ denote the current dark-energy, matter, and radiation densities of the Universe, respectively. 
For numerical values of all cosmological parameters, we use the latest Planck 2018 measurements~\cite{Aghanim:2018eyx}.

\begin{figure}
	\includegraphics[angle=0.0,width=0.48\textwidth]{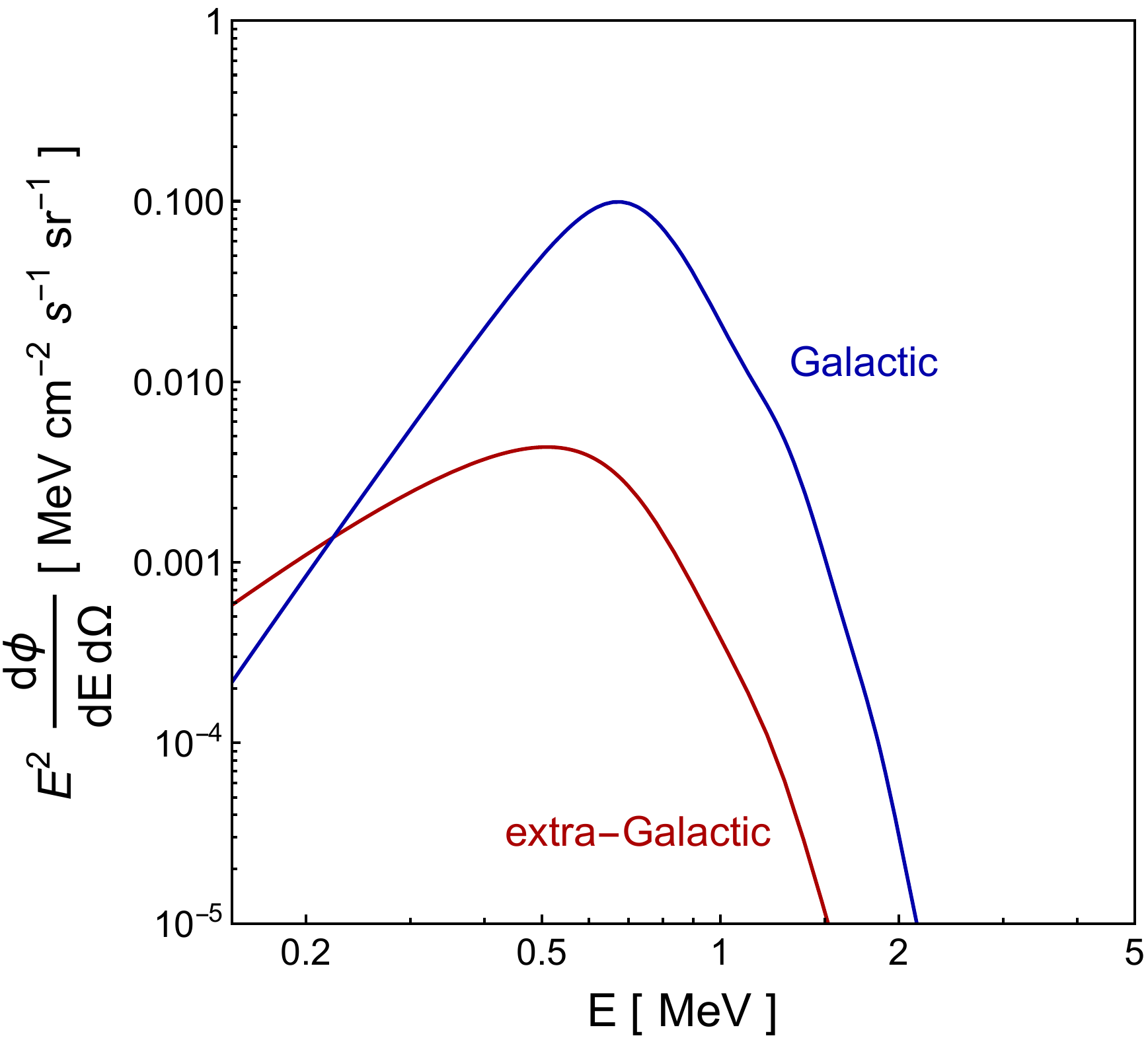}
	\caption{The Galactic and extra-Galactic photon contributions from Hawking evaporation off a non-rotating PBHs of mass $10^{17}$\,g. It is assumed that PBHs make up the entirety of DM and follow an NFW density profile.  The blue and red lines correspond to the Galactic and extra-Galactic contributions in the region of interest ($\abs{l} \leq 5 \deg$ and $\abs{b} \leq 5 \deg$) respectively.}
	\label{fig: Gal and EG}
\end{figure}

\begin{figure}
	\centering
	\includegraphics[angle=0.0,width=0.49\textwidth]{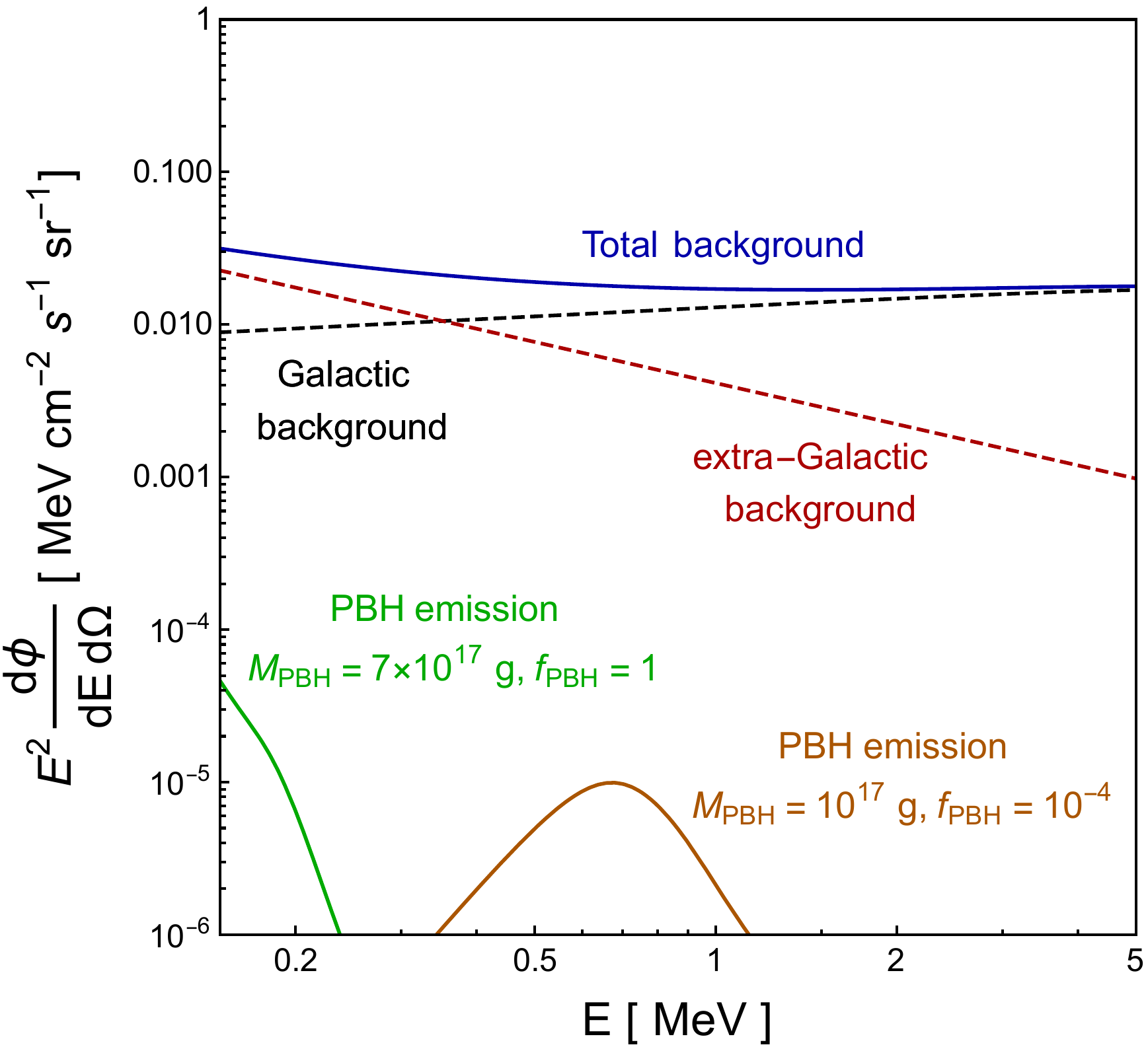}
	\caption{Galactic and extra-Galactic astrophysical backgrounds are shown as a function of the emitted photon energy. Dashed black line corresponds to the Galactic background which is adapted from Ref.~\cite{Bartels:2017dpb}. Dashed red line corresponds to the extra-Galactic  background which is a single power law fit to the Cosmic X-ray background measurements. Total background, sum of the Galactic and extra-Galactic backgrounds, is shown by the solid blue line. Evaporation signals from non-rotating PBHs of mass $10^{17}$ g with dark matter fraction of $10^{-4}$ and a non-rotating PBH of mass $7 \times 10^{17}$ g with dark matter fraction of unity are shown for comparison.}
		\label{fig: bkg}
\end{figure}

 In addition to a monochromatic mass distribution for PBHs, we also consider a log-normal mass distribution, as predicted by various inflationary models
\begin{equation}
\frac{dN_{\rm{PBH}}}{dM_{\rm{PBH}}}= \frac{1}{\sqrt{2\pi}\sigma M_{\rm{PBH}}}\, \exp\left[- \dfrac{ {\rm ln}^2 \left(M_{\rm PBH}/\mu_{\rm PBH}\right)}{2 \sigma^2} \right]\,,
\label{eq: log-normal mass distribution}
\end{equation} 
where $\mu_{\rm PBH}$ and $\sigma$ are the mean mass and width of the distribution. For an extended mass distribution of PBHs, the (extra-)Galactic contribution to the differential flux is
\begin{equation}
\frac{d\phi_{\rm gal,eg}} {dE}\Bigr\rvert_{\rm{ext}}=   \int dM_{\rm{PBH}}\,\frac{dN_{\rm{PBH}}}{dM_{\rm{PBH}}}\,\frac{d\phi_{\rm gal,eg}} {dE}\Bigr\rvert_{\rm{mono}}\,.
\label{eq: Galactic photon contribution extended}
\end{equation}

For non-rotating PBHs, the mass integral runs from $M_{\rm{min}} = 5 \times 10^{14}$ g to $M_{\rm{max}} =\infty$. For PBHs approaching to their maximal rotation, the mass integral runs from $M_{\rm{min}} = 7 \times 10^{14}$ g  to $M_{\rm{max}} =\infty$, as the maximal rotation increases the  minimum evaporation mass $M_{\rm{min}}$ by enhancing the Hawking emission rate. Note that, the minimum PBH mass only matters for extended PBH distributions with low averages.

Fig.\,\ref{fig: Gal and EG} shows the Galactic and extra-Galactic contributions to the total evaporation flux from PBHs of mass $10^{17}$ g in Galactic Center, defined to have: $\abs{l} \leq 5 \deg$ and $\abs{b} \leq 5 \deg$.  
Since this region of interest resides in a DM-dominated environment, the extra-Galactic contribution to the evaporation signal is always subdominant. Galactic emission peaks at around $\sim 0.6$  MeV as the temperature of a $10^{17}$ g PBH is 0.1 MeV.  The extra-Galactic signal peaks at a slightly lower energy as it is redshifted.

\begin{figure}
	\includegraphics[angle=0.0,width=0.48\textwidth]{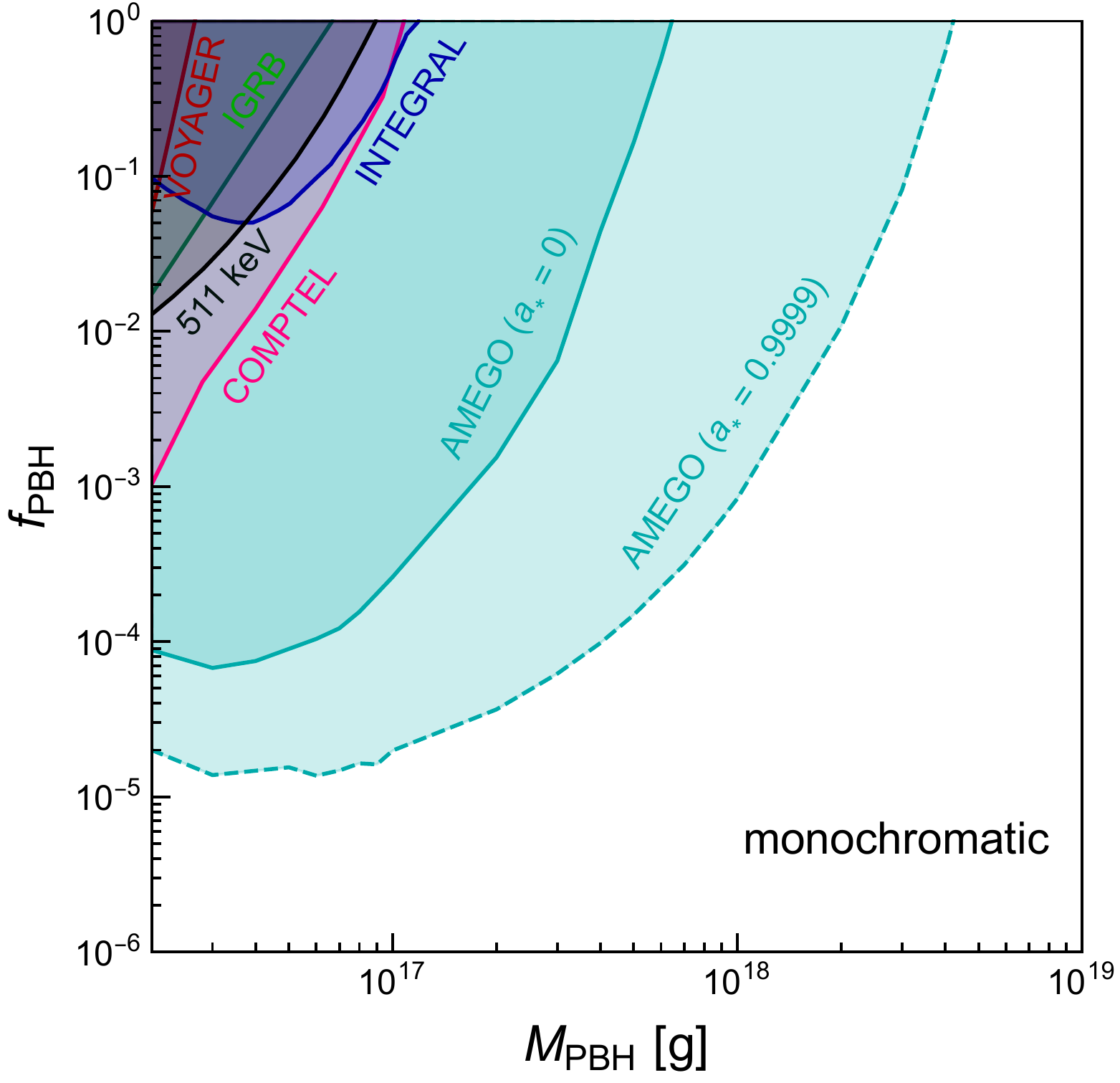}
	\caption{Projected upper limit (95\% C.L.) on the DM fraction of PBHs, $f_{\rm{PBH}}$, from near future MeV telescope AMEGO. The plot corresponds to a monochromatic mass distribution of PBHs. Results for non-rotating PBHs $(a_* = 0)$ and PBHs approaching to their maximal spin $(a_* = 0.9999)$ are shown by the solid line and the dashed line respectively.  The constraints are derived by considering an NFW density profile of the ultra-light PBHs. The existing exclusions on ultra-light non-spinning PBHs from Voyager-1 measurement of positron flux (shaded red)~\cite{Boudaud:2018hqb}, extra-Galactic gamma ray emission(shaded green)~\cite{Carr:2009jm,Ballesteros:2019exr,Arbey:2019vqx}, SPI/INTEGRAL  511 keV emission line with 1.5 kpc positron annihilation region \& isothermal DM profile (shaded black)~\cite{DeRocco:2019fjq,Laha:2019ssq,Dasgupta:2019cae} and INTEGRAL, COMPTEL Galactic Center MeV flux (shaded blue, shaded magenta)~\cite{Laha:2020ivk,Coogan:2020tuf} are also shown for comparison.  For reference, there are no existing exclusion limits to the right of the plot until $M_{\rm{PBH}} \sim 10^{23}$ g~\cite{Niikura:2017zjd,Smyth:2019whb,Montero-Camacho:2019jte,Katz:2018zrn}.	
}
	\label{fig: photon limits}
\end{figure}

\begin{figure}
	\includegraphics[angle=0.0,width=0.47\textwidth]{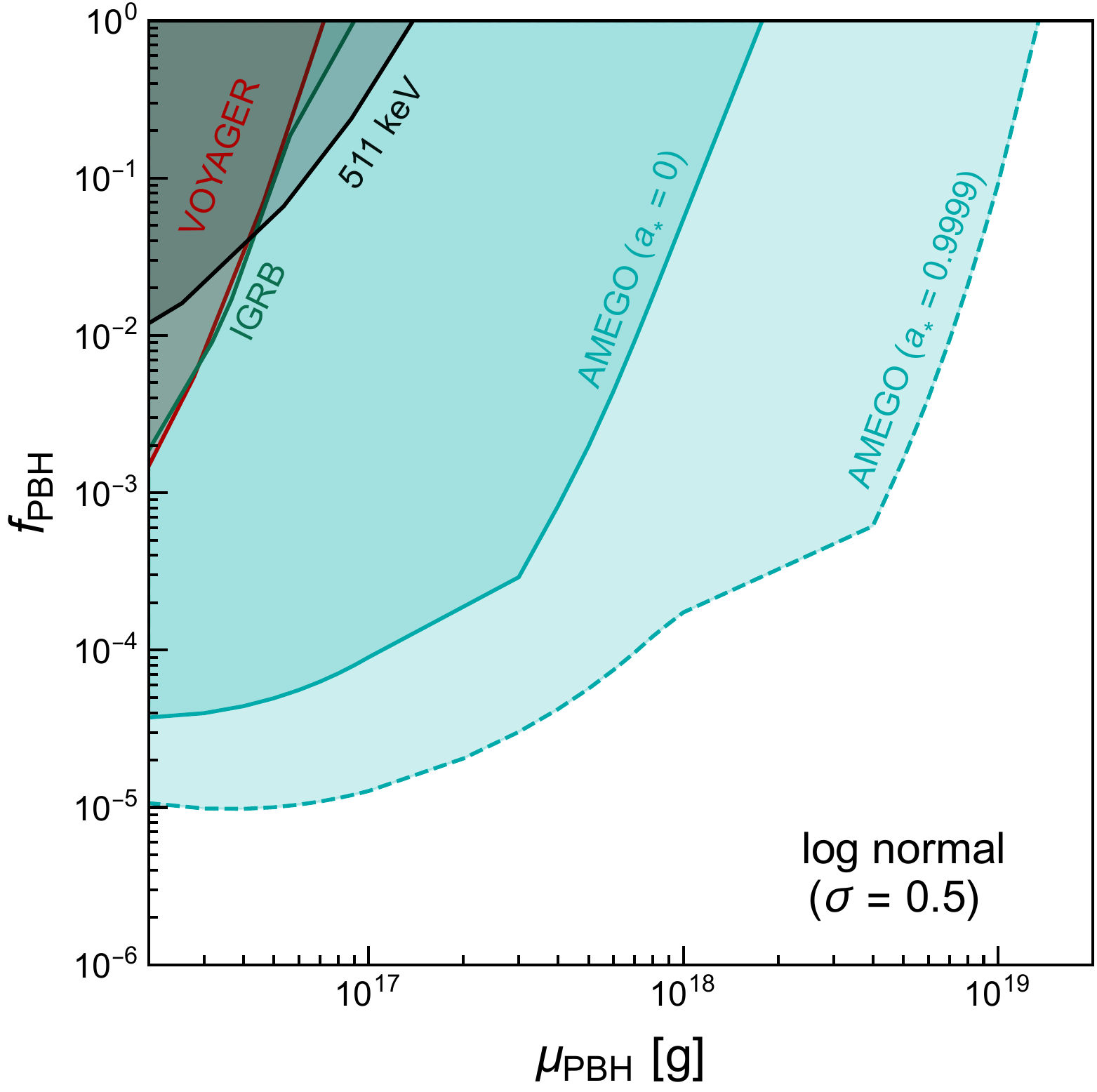}
	\caption{Projected upper limit (95\% C.L.) on the DM fraction of PBHs, $f_{\rm{PBH}}$, from near future MeV telescope AMEGO. Log-normal  mass distribution with a width  $\sigma = 0.5$ is considered in this plot.  Results for non-rotating PBHs $(a_* = 0)$ and maximally rotating PBHs $(a_* = 0.9999)$ are shown by the solid line and the dashed line respectively.  The constraints are derived by considering an NFW density profile of the ultra-light PBHs. The existing constraints on ultra-light non-spinning PBHs from Voyager-1 measurement of positron flux (shaded red)~\cite{Boudaud:2018hqb}, extra-Galactic gamma ray emission(shaded green)~\cite{Carr:2009jm,Ballesteros:2019exr,Arbey:2019vqx}, and SPI/INTEGRAL  511 keV emission line with 1.5 kpc positron annihilation region \& isothermal DM profile (shaded black)~\cite{DeRocco:2019fjq,Laha:2019ssq,Dasgupta:2019cae} are also shown for comparison.}
	\label{fig: photon limits extended}
\end{figure}

\begin{figure*}
	\centering
	\includegraphics[angle=0.0,width=1.0\textwidth]{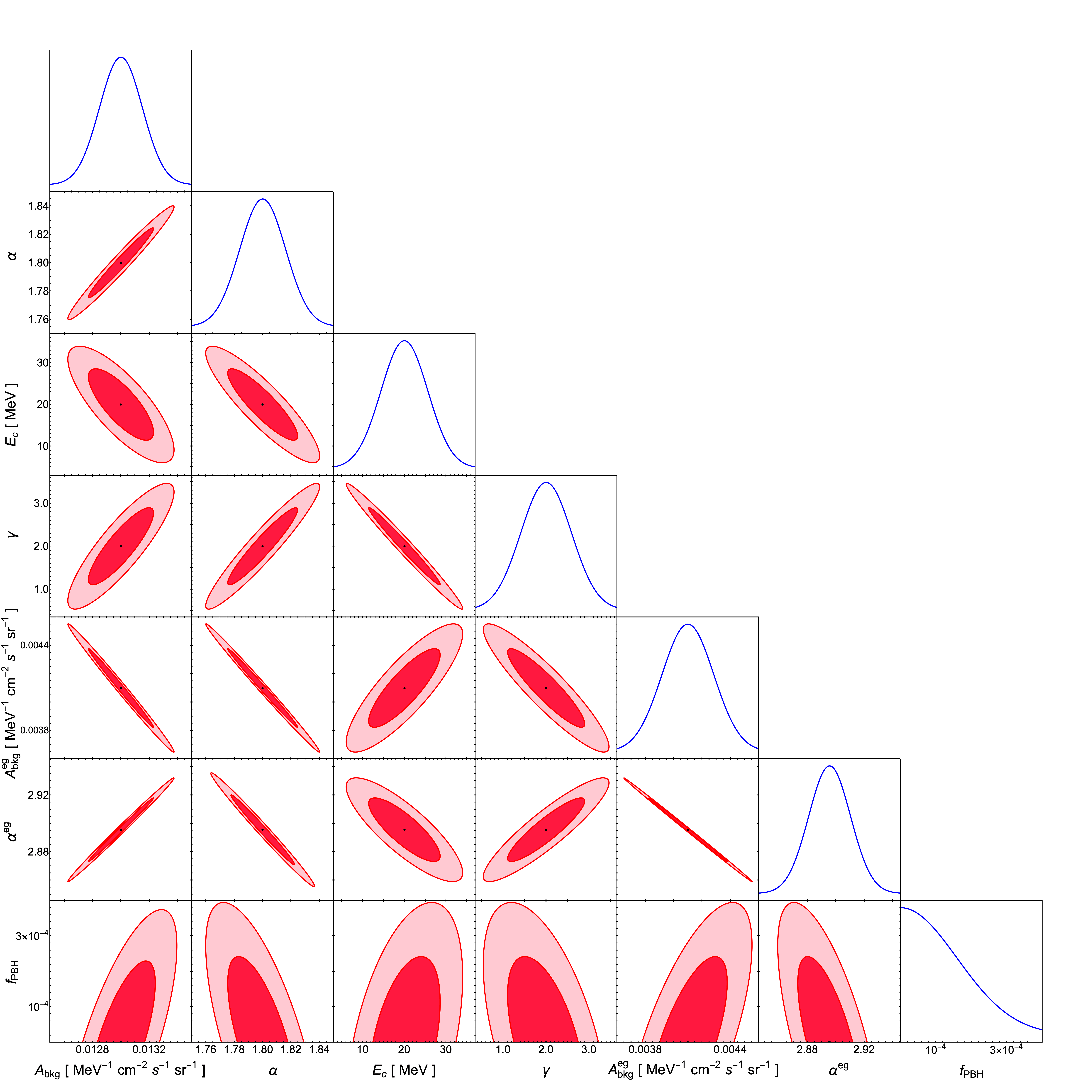}
	\caption{
	{
	Confidence ellipses at 68.3\% C.L. (1-$\sigma$, dark red) and 95\% C.L. (2-$\sigma$, light red)  for all background and signal parameters.  A larger correlation between parameters appears as a more tilted confidence ellipse. 
	For this corner plot we have assumed non-rotating and uncharged PBHs with a monochromatic mass distribution centered at $10^{17}$\,g, following a NFW density profile.  
	Black dots represent best-fit (i.e., our chosen fiducial) values of the background parameters, and we assume a fiducial $f_{\rm PBH}=0$. 
	The predicted posteriors for all parameters are shown by the Gaussian curves in blue. }
}
	\label{fig: confidence limits}
\end{figure*}

Of course, PBHs are not the only possible source of gamma-rays in the cosmos. 
In particular, there are well-known astrophysical backgrounds, which we ought to marginalize over to unearth a possible PBH signal.
Fig.\,\ref{fig: bkg} shows the Galactic and extra-Galactic astrophysical backgrounds used in this analysis. We have adapted the Galactic astrophysical background $\phi^{\rm{bkg}}_{\rm{gal}}$ from Ref.~\cite{Bartels:2017dpb}:
\begin{align}
 \phi^{\rm{bkg}}_{\rm{gal}} (E) = A_{\rm{bkg}} \left(\frac{E}{1\,\rm{MeV}}\right)^{-\alpha} \,\exp\left[{-\left(\frac{E}{E_c}\right)^\gamma}\right],
 \label{eq:gal bkg}
  \end{align}
 in units of $\rm{MeV}^{-1}\,cm^{-2}\,s^{-1}sr^{-1}$,
 which contains four parameters:  an amplitude $(A_{\rm{bkg}})$, power-law index $(\alpha)$, exponential cut-off energy $(E_{c})$, and the index within the exponential $(\gamma)$. 
 Their best fit values are  $A_{\rm{bkg}} = 0.013  \,{\rm{MeV}^{-1}\,cm^{-2}\,s^{-1}sr^{-1}},\, \alpha=1.8,\, E_c = 20 \,{\rm MeV},\, $and $ \gamma= 2$, respectively.
We have checked that this formula provides an adequate fit to the data obtained by COMPTEL\,\cite{1994A&A...292...82S,Strong:1998ck,Strong:2011pa}.  For the extra-Galactic background, $\phi^{\rm{bkg}}_{\rm{eg}}$, we have considered a single power law which fits the cosmic X-ray background spectrum measured by various experiments~\cite{1997AIPC..410.1223W,1975Ap&SS..32L...1F,Gruber:1999yr,1997ApJ...475..361K,Weidenspointner:2000aq} in the energy range 150 keV to 5 MeV~\cite{Ballesteros:2019exr}:
\begin{align}
\phi^{\rm{bkg}}_{\rm{eg}} (E) = A^{\rm{eg}}_{\rm{bkg}}\,\left(\frac{E}{1\,\rm{MeV}}\right)^{-\alpha^{\rm{eg}}}\,
 \label{eq:egal bkg}
\end{align}
also in $\rm{MeV}^{-1}\,cm^{-2}\,s^{-1}sr^{-1}$. 
Our power-law model for the extra-Galactic background contains two parameters, its amplitude $(A^{\rm{eg}}_{\rm{bkg}})$, and the power law index $(\alpha^{\rm{eg}})$, with best-fit values of $\,A^{\rm{eg}}_{\rm{bkg}} = 0.004135 \,\rm{MeV}^{-1}\,cm^{-2}\,s^{-1}sr^{-1}, $ and $ \alpha^{\rm{eg}}=2.8956$.

We consider photons in the energy range $0.15-5$ MeV for this analysis. The lower end of the energy range is determined by the sensitivity of AMEGO, whereas, the higher end of the energy range is determined by the evaporation signal. For PBHs of mass $2 \times 10^{16}$ g (minimum mass considered for this analysis), evaporation signal peaks at around 3 MeV, and falls off exponentially with increase in photon energy.
Moreover, the single power-law fit to the extra-Galactic background in Eq.(\ref{eq:egal bkg}) is valid only up to $\sim 5$ MeV~\cite{Ballesteros:2019exr}. 
  
  We have applied Fisher forecasting~\cite{Edwards:2017mnf,Edwards:2017kqw,doornhein2018uses} with marginalization over all astrophysical background parameters to compute the projected upper limits at 95\% C.L. The exclusion limits are derived by assuming no evaporation signal is present in the data. The Fisher information matrix $(\mathcal{F})$ is a $N \times N$ matrix, where $N$ denotes the total number of parameters $\vv{p}= \{p_1,p_2,...,p_N\}$ and is defined as~\cite{Bartels:2017dpb}
  \begin{equation}
  \mathcal{F}_{ij} = \int_{E}\int_{\Omega}\frac{\partial_i\phi (E, \Omega) \,\partial_j\phi (E, \Omega)}{\phi (E, \Omega)} \, T_{\rm{obs}} \, A_{\rm{eff}} (E) \, d\Omega \, dE\,,
  \end{equation}
  where $\phi(E,\Omega) = \left( \phi_{\rm{gal}}+ \phi_{\rm{eg}}+ \phi^{\rm{bkg}}_{\rm{gal}} +\phi^{\rm{bkg}}_{\rm{eg}} \right)$ is the total flux, $T_{\rm{obs}}$ is the observation time, and $A_{\rm{eff}} (E)$ is the effective area. 
  We conservatively ignore the extra-Galactic PBH emission, as it is subdominant in our region of interest. 
  The effective area for AMEGO is adapted from Ref.~\cite{2019BAAS...51g.245M}\footnote{\href{https://asd.gsfc.nasa.gov/amego/technical.html}{https://asd.gsfc.nasa.gov/amego/technical.html}}  and a uniform sky coverage $T_{\rm{obs}}$ of 1 year is considered for this analysis. We use a sufficiently dense binning in order to capture all the spectral variations in the Fisher information matrix.  We have considered 2000 logarithmically spaced bins between the energy interval of $0.15-5$ MeV.
  
   For this work, there is only one signal parameter, the fraction $f_{\rm{PBH}}$ of the DM that is composed of PBHs for each mass we study, plus the 6 astrophysical parameters introduced in Eqs.~\eqref{eq:gal bkg} and \eqref{eq:egal bkg}.
   Hence, the Fisher information matrix $(\mathcal{F})$  used in our analysis is a $7 \times 7$ symmetric matrix. The projected upper limit on the signal parameter $f_{\rm{PBH}}$ at 95\% C.L. is~\cite{Bartels:2017dpb}
\begin{equation}
f^{\rm{UL}}_{\rm{PBH}} = 1.645\,\sqrt{\,\left(\mathcal{F}^{-1}\right)_{11}}\,.
\end{equation}

  Because of the relatively large region of interest, our results are almost insensitive to different choices of DM density profiles. For this work, we assume that the density distribution of ultra-light PBHs in MW halo follows a Navarro-Frenk-White (NFW) profile~\cite{Navarro:1996gj}. However, we have tested our results with other density profiles such as with a cored isothermal profile~\cite{Ng:2013xha} and with a cored NFW profile with a core radii of 2 kpc~\cite{Laha:2020ivk}. We find that due to the different choices of DM density profiles, our results alter by as far as a factor of two (degrades by a factor of $\sim$ 1.69 for a cored isothermal profile and by a factor of $\sim$ 1.55 for a cored NFW profile with a core radii of 2 kpc). We have also checked that a somewhat larger region of interest around the Galactic Center, say $\abs{l} \leq 30 \deg$ and $\abs{b} \leq 10 \deg$,  increases the Hawking evaporated photons as well as background photons by a factor of  6.81 and 12 respectively, indicating a putative improvement of the projections by a factor of $(S/\sqrt{N}) \sim 1.96$ (which, however, may be reduced by marginalization with the Fisher matrix).
   
   Fig.\,\ref{fig: photon limits} shows the projected upper limits (at 95\% C.L.) on DM fraction of PBHs, $f_{\rm PBH}$, that can be derived from future AMEGO observations by assuming that no DM signal is present in the data. Monochromatic mass distribution of the PBHs is assumed in this plot. The solid/\,dashed lines correspond to non-rotating $(a_*= 0)$/approaching maximal rotation $(a_*= 0.9999)$ PBHs. Note that, we take the maximum value of spin as 0.9999 because {\tt BlackHawk} can not go beyond that.
    The limits are derived by assuming an NFW density profile of the PBHs. As the PBHs become maximally rotating, their temperatures as well as the effective energy of the emitted photons fall off rapidly, and as a result maximally rotating PBHs probe higher mass window than their non-spinning counterparts. Because of the lower energy reach and larger effective area, AMEGO is able to probe into asteroid-mass windows compared to the previous gamma-ray observatories such as INTEGRAL\,\cite{Winkler:2003nn}, Fermi\,\cite{Atwood:2009ez}, and CRGO\,\cite{1993ApJS...86..657S} for both non-rotating and maximally rotating PBHs. The projected upper limit from AMEGO excludes non-rotating (maximally rotating, $a_* = 0.9999$) PBHs as the sole component of DM upto $7 \times 10^{17}$ g ($4 \times 10^{18}$ g).  The kinks in the exclusion limits for both non-rotating and maximally rotating PBHs are due to finite number of mass point samplings.  Our exclusion limits start from $2 \times 10^{16}$\,g as lighter PBHs mostly evaporate to higher-energy photons, outside of our considered energy range. Quantitatively, for a non-rotating PBH of mass $10^{16}$\,g ($2 \times 10^{16}$\,g), $\sim 30\%$ ($\sim 97\%$) of the evaporation spectrum resides in our considered energy interval, explaining the choice of $2\times 10^{16}$\,g as the minimum PBH mass for this analysis.

 Fig.\,\ref{fig: photon limits extended} shows the projected upper limits (95\% C.L.) on DM fraction of PBHs, $f_{\rm PBH}$, that can be derived from future MeV telescope AMEGO by assuming no signal present in the data for an extended mass distribution.  Log-normal mass distribution of PBHs, a motivated scenario from several inflationary models, with a width of $\sigma = 0.5$ is considered to derive the exclusion limits. The density profile of PBHs are assumed to be NFW, however, the result degrades by at most a factor of two for cored density profiles. The solid (dashed) lines correspond to non-rotating (maximally rotating) PBHs. For this particular mass distribution, our projections exclude upto $\sim$ $2 \times 10^{18}$ g ($\sim 10^{19}$ g) for non-rotating (approaching maximal rotation $a_* = 0.9999$) PBHs. Similar to the monochromatic mass distributions, here also, AMEGO probes better than other proposed MeV telescopes because of its lower energy reach and larger effective area.  Similar to Fig.\,\ref{fig: photon limits}, here also, the kinks in the exclusion limits at around $\ 3 \times 10^{17}$ g for non-rotating PBHs, and at $\sim 3 \times 10^{18}$ g for maximally rotating PBHs are due to finite number of mass point samplings.

 Fig.\,\ref{fig: confidence limits} shows the confidence ellipses at 68.3\% C.L.  and 95\% C.L. for all signal and background parameters. Non-rotating PBHs with a monochromatic mass distribution centered at $10^{17}$ g and an NFW density profile is assumed for this figure. The confidence ellipses show degeneracies among all of the parameters and the parameters of the ellipses are computed from~\cite{Coe:2009xf}. For example, amplitude of the Galactic background $(A_{\rm{bkg}})$, amplitude of the extra-Galactic background $(A^{\rm{eg}}_{\rm{bkg}})$, and the exponential cutoff energy for the Galactic background $(E_{\rm{cut}})$ are correlated with the signal parameter, DM fraction of PBHs $(f_{\rm{PBH}})$. However, power law index of the Galactic background $(\alpha)$, power law index of the extra-Galactic background $(\alpha^{\rm{eg}})$, and index of the exponential cutoff energy in the Galactic background $(\gamma)$ are anti-correlated with $f_{\rm{PBH}}$. From the confidence ellipses, it is also evident that the correlation coefficient $r_{ij}(=\mathcal F^{-1}_{ij}/\sqrt{\mathcal F^{-1}_{ii} \mathcal F^{-1}_{jj}})$ between $A^{\rm{eg}}_{\rm{bkg}}$ and $f_{\rm{PBH}}$ $(r=0.697)$ is much stronger than the correlation between $E_{\rm{cut}}$ and $f_{\rm{PBH}}$ $(r=0.468)$. 
  In Fig.\,\ref{fig: confidence limits}, we also show the best fit values of all background parameters as well as their corresponding error bars by the mean and variance of the blue Gaussian curves.
 
\section{Summary and Conclusions}
 
PBHs in the asteroid-mass range, $\sim$ $10^{17}$ -- $10^{23}$ g, can make up the entire DM density and it is very important to conclusively probe these candidates.  We propose a strategy to decisively probe a part of this parameter space.  At the lower end of this mass range, PBHs with masses $\sim$ 10$^{17}$ g -- 10$^{18}$ g have Hawking temperatures in the range of 0.01 MeV to 0.1 MeV, implying that substantial evaporated photons are produced by them around these energy scales.  Near-future soft gamma-ray telescopes like AMEGO, with its large effective area and improved background rejection capabilities, can search for these photons and investigate this hard-to-probe parameter space.  The most efficient search strategy involves observations of the region around the Galactic Center.  We include the Galactic astrophysical background produced by cosmic-rays and the measured extra-galactic gamma-ray background in our projected search strategy.  Our projections show that AMEGO can exclude non-rotating PBHs as the sole component of DM upto $\sim$ $7 \times 10^{17}$ g.  We demonstrate that maximal rotation as well as extended mass distribution of the PBHs allow us to explore larger ranges of PBH masses. We also predict that the projected exclusions on PBH DM in the mass range $\sim$ $10^{16}$ $-$ $10^{17}$ g will be much stronger than the existing limits. The projections presented in this work are robust to the different choices of DM density profiles.  At higher PBH masses in this range, the Hawking radiation flux gets smaller and thus much larger instruments need to be built in order to detect the evaporation signature.  In the absence of much larger telescopes, other techniques need to be developed in order to probe the complete parameter space of asteroid-mass PBHs.

\section{ACKNOWLEDGMENTS}
 
We thank Basudeb Dasgupta and Thomas Edwards for discussions and useful suggestions.  We especially thank Tracy R. Slatyer for early discussions which led to this work. JBM is supported by a Clay Fellowship at the Smithsonian Astrophysical Observatory.

\bibliographystyle{JHEP}
 \bibliography{ref}

\providecommand{\href}[2]{#2}\begingroup\raggedright\begin{thebibliography}{100}

\bibitem{Aghanim:2018eyx}
{\scshape Planck} collaboration, N.~Aghanim et~al., \emph{{Planck 2018 results.
  VI. Cosmological parameters}},
  \href{https://doi.org/10.1051/0004-6361/201833910}{\emph{Astron. Astrophys.}
  {\bfseries 641} (2020) A6}
  [\href{https://arxiv.org/abs/1807.06209}{{\ttfamily 1807.06209}}].

\bibitem{Pardo:2020epc}
K.~Pardo and D.~N. Spergel, \emph{{What is the price of abandoning dark matter?
  Cosmological constraints on alternative gravity theories}},
  \href{https://doi.org/10.1103/PhysRevLett.125.211101}{\emph{Phys. Rev. Lett.}
  {\bfseries 125} (2020) 211101}
  [\href{https://arxiv.org/abs/2007.00555}{{\ttfamily 2007.00555}}].

\bibitem{Strigari:2013iaa}
L.~E. Strigari, \emph{{Galactic Searches for Dark Matter}},
  \href{https://doi.org/10.1016/j.physrep.2013.05.004}{\emph{Phys. Rept.}
  {\bfseries 531} (2013) 1} [\href{https://arxiv.org/abs/1211.7090}{{\ttfamily
  1211.7090}}].

\bibitem{Slatyer:2017sev}
T.~R. Slatyer, \emph{{Indirect Detection of Dark Matter}},  in
  \emph{{Theoretical Advanced Study Institute in Elementary Particle Physics}:
  {Anticipating the Next Discoveries in Particle Physics}}, pp.~297--353, 2018,
  \href{https://arxiv.org/abs/1710.05137}{{\ttfamily 1710.05137}},
  \href{https://doi.org/10.1142/9789813233348_0005}{DOI}.

\bibitem{Lin:2019uvt}
T.~Lin, \emph{{Dark matter models and direct detection}},
  \href{https://doi.org/10.22323/1.333.0009}{\emph{PoS} {\bfseries 333} (2019)
  009} [\href{https://arxiv.org/abs/1904.07915}{{\ttfamily 1904.07915}}].

\bibitem{Boveia:2018yeb}
A.~Boveia and C.~Doglioni, \emph{{Dark Matter Searches at Colliders}},
  \href{https://doi.org/10.1146/annurev-nucl-101917-021008}{\emph{Ann. Rev.
  Nucl. Part. Sci.} {\bfseries 68} (2018) 429}
  [\href{https://arxiv.org/abs/1810.12238}{{\ttfamily 1810.12238}}].

\bibitem{1966AZh....43..758Z}
Y.~B. {Zel'dovich} and I.~D. {Novikov}, \emph{{The Hypothesis of Cores Retarded
  during Expansion and the Hot Cosmological Model}}, {\emph{Astronomicheskii
  Zhurnal} {\bfseries 43} (1966) 758}.

\bibitem{Hawking:1971ei}
S.~Hawking, \emph{{Gravitationally collapsed objects of very low mass}},
  \href{https://doi.org/10.1093/mnras/152.1.75}{\emph{Mon. Not. Roy. Astron.
  Soc.} {\bfseries 152} (1971) 75}.

\bibitem{Carr:1974nx}
B.~J. Carr and S.~W. Hawking, \emph{{Black holes in the early Universe}},
  \href{https://doi.org/10.1093/mnras/168.2.399}{\emph{Mon. Not. Roy. Astron.
  Soc.} {\bfseries 168} (1974) 399}.

\bibitem{Chapline:1975ojl}
G.~F. Chapline, \emph{{Cosmological effects of primordial black holes}},
  \href{https://doi.org/10.1038/253251a0}{\emph{Nature} {\bfseries 253} (1975)
  251}.

\bibitem{Katz:2018zrn}
A.~Katz, J.~Kopp, S.~Sibiryakov and W.~Xue, \emph{{Femtolensing by Dark Matter
  Revisited}}, \href{https://doi.org/10.1088/1475-7516/2018/12/005}{\emph{JCAP}
  {\bfseries 1812} (2018) 005}
  [\href{https://arxiv.org/abs/1807.11495}{{\ttfamily 1807.11495}}].

\bibitem{Montero-Camacho:2019jte}
P.~Montero-Camacho, X.~Fang, G.~Vasquez, M.~Silva and C.~M. Hirata,
  \emph{{Revisiting constraints on asteroid-mass primordial black holes as dark
  matter candidates}},
  \href{https://doi.org/10.1088/1475-7516/2019/08/031}{\emph{JCAP} {\bfseries
  1908} (2019) 031} [\href{https://arxiv.org/abs/1906.05950}{{\ttfamily
  1906.05950}}].

\bibitem{Smyth:2019whb}
N.~Smyth, S.~Profumo, S.~English, T.~Jeltema, K.~McKinnon and P.~Guhathakurta,
  \emph{{Updated Constraints on Asteroid-Mass Primordial Black Holes as Dark
  Matter}}, \href{https://doi.org/10.1103/PhysRevD.101.063005}{\emph{Phys. Rev.
  D} {\bfseries 101} (2020) 063005}
  [\href{https://arxiv.org/abs/1910.01285}{{\ttfamily 1910.01285}}].

\bibitem{Dasgupta:2019cae}
B.~Dasgupta, R.~Laha and A.~Ray, \emph{{Neutrino and positron constraints on
  spinning primordial black hole dark matter}},
  \href{https://doi.org/10.1103/PhysRevLett.125.101101}{\emph{Phys. Rev. Lett.}
  {\bfseries 125} (2020) 101101}
  [\href{https://arxiv.org/abs/1912.01014}{{\ttfamily 1912.01014}}].

\bibitem{Carr:2020gox}
B.~Carr, K.~Kohri, Y.~Sendouda and J.~Yokoyama, \emph{{Constraints on
  Primordial Black Holes}},  \href{https://arxiv.org/abs/2002.12778}{{\ttfamily
  2002.12778}}.

\bibitem{Green:2020jor}
A.~M. Green and B.~J. Kavanagh, \emph{{Primordial Black Holes as a dark matter
  candidate}}, \href{https://doi.org/10.1088/1361-6471/abc534}{\emph{J. Phys.
  G} {\bfseries 48} (2021) 4}
  [\href{https://arxiv.org/abs/2007.10722}{{\ttfamily 2007.10722}}].

\bibitem{Carr:2020xqk}
B.~Carr and F.~Kuhnel, \emph{{Primordial Black Holes as Dark Matter: Recent
  Developments}},
  \href{https://doi.org/10.1146/annurev-nucl-050520-125911}{\emph{Ann. Rev.
  Nucl. Part. Sci.} {\bfseries 70} (2020) 355}
  [\href{https://arxiv.org/abs/2006.02838}{{\ttfamily 2006.02838}}].

\bibitem{Abbott:2016blz}
{\scshape LIGO Scientific, Virgo} collaboration, B.~P. Abbott et~al.,
  \emph{{Observation of Gravitational Waves from a Binary Black Hole Merger}},
  \href{https://doi.org/10.1103/PhysRevLett.116.061102}{\emph{Phys. Rev. Lett.}
  {\bfseries 116} (2016) 061102}
  [\href{https://arxiv.org/abs/1602.03837}{{\ttfamily 1602.03837}}].

\bibitem{Bird:2016dcv}
S.~Bird, I.~Cholis, J.~B. Muñoz, Y.~Ali-Haimoud, M.~Kamionkowski, E.~D. Kovetz
  et~al., \emph{{Did LIGO detect dark matter?}},
  \href{https://doi.org/10.1103/PhysRevLett.116.201301}{\emph{Phys. Rev. Lett.}
  {\bfseries 116} (2016) 201301}
  [\href{https://arxiv.org/abs/1603.00464}{{\ttfamily 1603.00464}}].

\bibitem{Clesse:2016vqa}
S.~Clesse and J.~Garcia-Bellido, \emph{{The clustering of massive Primordial
  Black Holes as Dark Matter: measuring their mass distribution with Advanced
  LIGO}}, \href{https://doi.org/10.1016/j.dark.2016.10.002}{\emph{Phys. Dark
  Univ.} {\bfseries 15} (2017) 142}
  [\href{https://arxiv.org/abs/1603.05234}{{\ttfamily 1603.05234}}].

\bibitem{Sasaki:2016jop}
M.~Sasaki, T.~Suyama, T.~Tanaka and S.~Yokoyama, \emph{{Primordial Black Hole
  Scenario for the Gravitational-Wave Event GW150914}},
  \href{https://doi.org/10.1103/PhysRevLett.121.059901,
  10.1103/PhysRevLett.117.061101}{\emph{Phys. Rev. Lett.} {\bfseries 117}
  (2016) 061101} [\href{https://arxiv.org/abs/1603.08338}{{\ttfamily
  1603.08338}}].

\bibitem{Arbey:2019vqx}
A.~Arbey, J.~Auffinger and J.~Silk, \emph{{Constraining primordial black hole
  masses with the isotropic gamma ray background}},
  \href{https://doi.org/10.1103/PhysRevD.101.023010}{\emph{Phys. Rev. D}
  {\bfseries 101} (2020) 023010}
  [\href{https://arxiv.org/abs/1906.04750}{{\ttfamily 1906.04750}}].

\bibitem{Clark:2016nst}
S.~Clark, B.~Dutta, Y.~Gao, L.~E. Strigari and S.~Watson, \emph{{Planck
  Constraint on Relic Primordial Black Holes}},
  \href{https://doi.org/10.1103/PhysRevD.95.083006}{\emph{Phys. Rev. D}
  {\bfseries 95} (2017) 083006}
  [\href{https://arxiv.org/abs/1612.07738}{{\ttfamily 1612.07738}}].

\bibitem{Wang:2016ana}
S.~Wang, Y.-F. Wang, Q.-G. Huang and T.~G.~F. Li, \emph{{Constraints on the
  Primordial Black Hole Abundance from the First Advanced LIGO Observation Run
  Using the Stochastic Gravitational-Wave Background}},
  \href{https://doi.org/10.1103/PhysRevLett.120.191102}{\emph{Phys. Rev. Lett.}
  {\bfseries 120} (2018) 191102}
  [\href{https://arxiv.org/abs/1610.08725}{{\ttfamily 1610.08725}}].

\bibitem{Poulter:2019ooo}
H.~Poulter, Y.~Ali-Haimoud, J.~Hamann, M.~White and A.~G. Williams, \emph{{CMB
  constraints on ultra-light primordial black holes with extended mass
  distributions}},  \href{https://arxiv.org/abs/1907.06485}{{\ttfamily
  1907.06485}}.

\bibitem{Boudaud:2018hqb}
M.~Boudaud and M.~Cirelli, \emph{{Voyager 1 e$^\pm$ Further Constrain
  Primordial Black Holes as Dark Matter}},
  \href{https://doi.org/10.1103/PhysRevLett.122.041104}{\emph{Phys. Rev. Lett.}
  {\bfseries 122} (2019) 041104}
  [\href{https://arxiv.org/abs/1807.03075}{{\ttfamily 1807.03075}}].

\bibitem{DeRocco:2019fjq}
W.~DeRocco and P.~W. Graham, \emph{{Constraining Primordial Black Hole
  Abundance with the Galactic 511 keV Line}},
  \href{https://doi.org/10.1103/PhysRevLett.123.251102}{\emph{Phys. Rev. Lett.}
  {\bfseries 123} (2019) 251102}
  [\href{https://arxiv.org/abs/1906.07740}{{\ttfamily 1906.07740}}].

\bibitem{Laha:2019ssq}
R.~Laha, \emph{{Primordial Black Holes as a Dark Matter Candidate Are Severely
  Constrained by the Galactic Center 511 keV $\gamma$ -Ray Line}},
  \href{https://doi.org/10.1103/PhysRevLett.123.251101}{\emph{Phys. Rev. Lett.}
  {\bfseries 123} (2019) 251101}
  [\href{https://arxiv.org/abs/1906.09994}{{\ttfamily 1906.09994}}].

\bibitem{Ballesteros:2019exr}
G.~Ballesteros, J.~Coronado-Bl\'azquez and D.~Gaggero, \emph{{X-ray and
  gamma-ray limits on the primordial black hole abundance from Hawking
  radiation}},
  \href{https://doi.org/10.1016/j.physletb.2020.135624}{\emph{Phys. Lett. B}
  {\bfseries 808} (2020) 135624}
  [\href{https://arxiv.org/abs/1906.10113}{{\ttfamily 1906.10113}}].

\bibitem{Laha:2020ivk}
R.~Laha, J.~B. Munoz and T.~R. Slatyer, \emph{{INTEGRAL constraints on
  primordial black holes and particle dark matter}},
  \href{https://doi.org/10.1103/PhysRevD.101.123514}{\emph{Phys. Rev. D}
  {\bfseries 101} (2020) 123514}
  [\href{https://arxiv.org/abs/2004.00627}{{\ttfamily 2004.00627}}].

\bibitem{Allsman:2000kg}
{\scshape Macho} collaboration, R.~Allsman et~al., \emph{{MACHO project limits
  on black hole dark matter in the 1-30 solar mass range}},
  \href{https://doi.org/10.1086/319636}{\emph{Astrophys. J. Lett.} {\bfseries
  550} (2001) L169} [\href{https://arxiv.org/abs/astro-ph/0011506}{{\ttfamily
  astro-ph/0011506}}].

\bibitem{Tisserand:2006zx}
{\scshape EROS-2} collaboration, P.~Tisserand et~al., \emph{{Limits on the
  Macho Content of the Galactic Halo from the EROS-2 Survey of the Magellanic
  Clouds}}, \href{https://doi.org/10.1051/0004-6361:20066017}{\emph{Astron.
  Astrophys.} {\bfseries 469} (2007) 387}
  [\href{https://arxiv.org/abs/astro-ph/0607207}{{\ttfamily
  astro-ph/0607207}}].

\bibitem{Niikura:2019kqi}
H.~Niikura, M.~Takada, S.~Yokoyama, T.~Sumi and S.~Masaki, \emph{{Constraints
  on Earth-mass primordial black holes from OGLE 5-year microlensing events}},
  \href{https://doi.org/10.1103/PhysRevD.99.083503}{\emph{Phys. Rev. D}
  {\bfseries 99} (2019) 083503}
  [\href{https://arxiv.org/abs/1901.07120}{{\ttfamily 1901.07120}}].

\bibitem{Oguri:2017ock}
M.~Oguri, J.~M. Diego, N.~Kaiser, P.~L. Kelly and T.~Broadhurst,
  \emph{{Understanding caustic crossings in giant arcs: characteristic scales,
  event rates, and constraints on compact dark matter}},
  \href{https://doi.org/10.1103/PhysRevD.97.023518}{\emph{Phys. Rev. D}
  {\bfseries 97} (2018) 023518}
  [\href{https://arxiv.org/abs/1710.00148}{{\ttfamily 1710.00148}}].

\bibitem{Zumalacarregui:2017qqd}
M.~Zumalacarregui and U.~Seljak, \emph{{Limits on stellar-mass compact objects
  as dark matter from gravitational lensing of type Ia supernovae}},
  \href{https://doi.org/10.1103/PhysRevLett.121.141101}{\emph{Phys. Rev. Lett.}
  {\bfseries 121} (2018) 141101}
  [\href{https://arxiv.org/abs/1712.02240}{{\ttfamily 1712.02240}}].

\bibitem{Authors:2019qbw}
{\scshape LIGO Scientific, Virgo} collaboration, B.~Abbott et~al.,
  \emph{{Search for Subsolar Mass Ultracompact Binaries in Advanced LIGO's
  Second Observing Run}},
  \href{https://doi.org/10.1103/PhysRevLett.123.161102}{\emph{Phys. Rev. Lett.}
  {\bfseries 123} (2019) 161102}
  [\href{https://arxiv.org/abs/1904.08976}{{\ttfamily 1904.08976}}].

\bibitem{Kavanagh:2018ggo}
B.~J. Kavanagh, D.~Gaggero and G.~Bertone, \emph{{Merger rate of a subdominant
  population of primordial black holes}},
  \href{https://doi.org/10.1103/PhysRevD.98.023536}{\emph{Phys. Rev. D}
  {\bfseries 98} (2018) 023536}
  [\href{https://arxiv.org/abs/1805.09034}{{\ttfamily 1805.09034}}].

\bibitem{Brandt:2016aco}
T.~D. Brandt, \emph{{Constraints on MACHO Dark Matter from Compact Stellar
  Systems in Ultra-Faint Dwarf Galaxies}},
  \href{https://doi.org/10.3847/2041-8205/824/2/L31}{\emph{Astrophys. J. Lett.}
  {\bfseries 824} (2016) L31}
  [\href{https://arxiv.org/abs/1605.03665}{{\ttfamily 1605.03665}}].

\bibitem{Koushiappas:2017chw}
S.~M. Koushiappas and A.~Loeb, \emph{{Dynamics of Dwarf Galaxies Disfavor
  Stellar-Mass Black Holes as Dark Matter}},
  \href{https://doi.org/10.1103/PhysRevLett.119.041102}{\emph{Phys. Rev. Lett.}
  {\bfseries 119} (2017) 041102}
  [\href{https://arxiv.org/abs/1704.01668}{{\ttfamily 1704.01668}}].

\bibitem{Monroy-Rodriguez:2014ula}
M.~A. Monroy-Rodriguez and C.~Allen, \emph{{The end of the MACHO era-
  revisited: new limits on MACHO masses from halo wide binaries}},
  \href{https://doi.org/10.1088/0004-637X/790/2/159}{\emph{Astrophys. J.}
  {\bfseries 790} (2014) 159}
  [\href{https://arxiv.org/abs/1406.5169}{{\ttfamily 1406.5169}}].

\bibitem{Serpico:2020ehh}
P.~D. Serpico, V.~Poulin, D.~Inman and K.~Kohri, \emph{{Cosmic microwave
  background bounds on primordial black holes including dark matter halo
  accretion}},
  \href{https://doi.org/10.1103/PhysRevResearch.2.023204}{\emph{Phys. Rev.
  Res.} {\bfseries 2} (2020) 023204}
  [\href{https://arxiv.org/abs/2002.10771}{{\ttfamily 2002.10771}}].

\bibitem{Hektor:2018qqw}
A.~Hektor, G.~Hutsi, L.~Marzola, M.~Raidal, V.~Vaskonen and H.~Veermae,
  \emph{{Constraining Primordial Black Holes with the EDGES 21-cm Absorption
  Signal}}, \href{https://doi.org/10.1103/PhysRevD.98.023503}{\emph{Phys. Rev.
  D} {\bfseries 98} (2018) 023503}
  [\href{https://arxiv.org/abs/1803.09697}{{\ttfamily 1803.09697}}].

\bibitem{Manshanden:2018tze}
J.~Manshanden, D.~Gaggero, G.~Bertone, R.~M. Connors and M.~Ricotti,
  \emph{{Multi-wavelength astronomical searches for primordial black holes}},
  \href{https://doi.org/10.1088/1475-7516/2019/06/026}{\emph{JCAP} {\bfseries
  06} (2019) 026} [\href{https://arxiv.org/abs/1812.07967}{{\ttfamily
  1812.07967}}].

\bibitem{Hektor:2018rul}
A.~Hektor, G.~Hutsi and M.~Raidal, \emph{{Constraints on primordial black hole
  dark matter from Galactic center X-ray observations}},
  \href{https://doi.org/10.1051/0004-6361/201833483}{\emph{Astron. Astrophys.}
  {\bfseries 618} (2018) A139}
  [\href{https://arxiv.org/abs/1805.06513}{{\ttfamily 1805.06513}}].

\bibitem{Raidal:2017mfl}
M.~Raidal, V.~Vaskonen and H.~Veerm\"ae, \emph{{Gravitational Waves from
  Primordial Black Hole Mergers}},
  \href{https://doi.org/10.1088/1475-7516/2017/09/037}{\emph{JCAP} {\bfseries
  09} (2017) 037} [\href{https://arxiv.org/abs/1707.01480}{{\ttfamily
  1707.01480}}].

\bibitem{Sammons:2020kyk}
M.~W. Sammons, J.-P. Macquart, R.~D. Ekers, R.~M. Shannon, H.~Cho, J.~X.
  Prochaska et~al., \emph{{First Constraints on Compact Dark Matter from Fast
  Radio Burst Microstructure}},
  \href{https://doi.org/10.3847/1538-4357/aba7bb}{\emph{Astrophys. J.}
  {\bfseries 900} (2020) 122}
  [\href{https://arxiv.org/abs/2002.12533}{{\ttfamily 2002.12533}}].

\bibitem{Lu:2020bmd}
P.~Lu, V.~Takhistov, G.~B. Gelmini, K.~Hayashi, Y.~Inoue and A.~Kusenko,
  \emph{{Constraining Primordial Black Holes with Dwarf Galaxy Heating}},
  \href{https://doi.org/10.3847/2041-8213/abdcb6}{\emph{Astrophys. J. Lett.}
  {\bfseries 908} (2021) L23}
  [\href{https://arxiv.org/abs/2007.02213}{{\ttfamily 2007.02213}}].

\bibitem{Nitz:2020bdb}
A.~H. Nitz and Y.-F. Wang, \emph{{Search for Gravitational Waves from
  High-Mass-Ratio Compact-Binary Mergers of Stellar Mass and Subsolar Mass
  Black Holes}},
  \href{https://doi.org/10.1103/PhysRevLett.126.021103}{\emph{Phys. Rev. Lett.}
  {\bfseries 126} (2021) 021103}
  [\href{https://arxiv.org/abs/2007.03583}{{\ttfamily 2007.03583}}].

\bibitem{Nitz:2021mzz}
A.~H. Nitz and Y.-F. Wang, \emph{{Search for gravitational waves from the
  coalescence of sub-solar mass and eccentric compact binaries}},
  \href{https://arxiv.org/abs/2102.00868}{{\ttfamily 2102.00868}}.

\bibitem{Laha:2020vhg}
R.~Laha, P.~Lu and V.~Takhistov, \emph{{Gas Heating from Spinning and
  Non-Spinning Evaporating Primordial Black Holes}},
  \href{https://arxiv.org/abs/2009.11837}{{\ttfamily 2009.11837}}.

\bibitem{2021MNRAS.504.5475K}
H.~{Kim}, \emph{{A constraint on light primordial black holes from the
  interstellar medium temperature}},
  \href{https://doi.org/10.1093/mnras/stab1222}{\emph{Monthly Notices of the
  Royal Astronomical Society} {\bfseries 504} (2021) 5475}
  [\href{https://arxiv.org/abs/2007.07739}{{\ttfamily 2007.07739}}].

\bibitem{Chan:2020zry}
M.~H. Chan and C.~M. Lee, \emph{{Constraining Primordial Black Hole Fraction at
  the Galactic Centre using radio observational data}},
  \href{https://doi.org/10.1093/mnras/staa1966}{\emph{Mon. Not. Roy. Astron.
  Soc.} {\bfseries 497} (2020) 1212}
  [\href{https://arxiv.org/abs/2007.05677}{{\ttfamily 2007.05677}}].

\bibitem{Dolgov:2020xzo}
A.~Dolgov, A.~Kuranov, N.~Mitichkin, S.~Porey, K.~Postnov, O.~Sazhina et~al.,
  \emph{{On mass distribution of coalescing black holes}},
  \href{https://doi.org/10.1088/1475-7516/2020/12/017}{\emph{JCAP} {\bfseries
  12} (2020) 017} [\href{https://arxiv.org/abs/2005.00892}{{\ttfamily
  2005.00892}}].

\bibitem{Wong:2020yig}
K.~W.~K. Wong, G.~Franciolini, V.~De~Luca, V.~Baibhav, E.~Berti, P.~Pani
  et~al., \emph{{Constraining the primordial black hole scenario with Bayesian
  inference and machine learning: the GWTC-2 gravitational wave catalog}},
  \href{https://doi.org/10.1103/PhysRevD.103.023026}{\emph{Phys. Rev. D}
  {\bfseries 103} (2021) 023026}
  [\href{https://arxiv.org/abs/2011.01865}{{\ttfamily 2011.01865}}].

\bibitem{Hutsi:2020sol}
G.~H\"utsi, M.~Raidal, V.~Vaskonen and H.~Veerm\"ae, \emph{{Two populations of
  LIGO-Virgo black holes}},
  \href{https://doi.org/10.1088/1475-7516/2021/03/068}{\emph{JCAP} {\bfseries
  03} (2021) 068} [\href{https://arxiv.org/abs/2012.02786}{{\ttfamily
  2012.02786}}].

\bibitem{Coogan:2020tuf}
A.~Coogan, L.~Morrison and S.~Profumo, \emph{{Direct Detection of Hawking
  Radiation from Asteroid-Mass Primordial Black Holes}},
  \href{https://doi.org/10.1103/PhysRevLett.126.171101}{\emph{Phys. Rev. Lett.}
  {\bfseries 126} (2021) 171101}
  [\href{https://arxiv.org/abs/2010.04797}{{\ttfamily 2010.04797}}].

\bibitem{Halder:2021jiv}
A.~Halder and M.~Pandey, \emph{{Investigating the Effect of PBH, Dark Matter --
  Baryon and Dark Matter -- Dark Energy Interaction on EDGES in 21cm Signal}},
  \href{https://arxiv.org/abs/2101.05228}{{\ttfamily 2101.05228}}.

\bibitem{Munoz:2016tmg}
J.~B. Munoz, E.~D. Kovetz, L.~Dai and M.~Kamionkowski, \emph{{Lensing of Fast
  Radio Bursts as a Probe of Compact Dark Matter}},
  \href{https://doi.org/10.1103/PhysRevLett.117.091301}{\emph{Phys. Rev. Lett.}
  {\bfseries 117} (2016) 091301}
  [\href{https://arxiv.org/abs/1605.00008}{{\ttfamily 1605.00008}}].

\bibitem{Laha:2018zav}
R.~Laha, \emph{{Lensing of fast radio bursts: Future constraints on primordial
  black hole density with an extended mass function and a new probe of exotic
  compact fermion and boson stars}},
  \href{https://doi.org/10.1103/PhysRevD.102.023016}{\emph{Phys. Rev. D}
  {\bfseries 102} (2020) 023016}
  [\href{https://arxiv.org/abs/1812.11810}{{\ttfamily 1812.11810}}].

\bibitem{Katz:2019qug}
A.~Katz, J.~Kopp, S.~Sibiryakov and W.~Xue, \emph{{Looking for MACHOs in the
  Spectra of Fast Radio Bursts}},
  \href{https://doi.org/10.1093/mnras/staa1497}{\emph{Mon. Not. Roy. Astron.
  Soc.} {\bfseries 496} (2020) 564}
  [\href{https://arxiv.org/abs/1912.07620}{{\ttfamily 1912.07620}}].

\bibitem{Jung:2017flg}
S.~Jung and C.~S. Shin, \emph{{Gravitational-Wave Fringes at LIGO: Detecting
  Compact Dark Matter by Gravitational Lensing}},
  \href{https://doi.org/10.1103/PhysRevLett.122.041103}{\emph{Phys. Rev. Lett.}
  {\bfseries 122} (2019) 041103}
  [\href{https://arxiv.org/abs/1712.01396}{{\ttfamily 1712.01396}}].

\bibitem{Kuhnel:2018mlr}
F.~Kuhnel, A.~Matas, G.~D. Starkman and K.~Freese, \emph{{Waves from the
  Centre: Probing PBH and other Macroscopic Dark Matter with LISA}},
  \href{https://doi.org/10.1140/epjc/s10052-020-8183-4}{\emph{Eur. Phys. J. C}
  {\bfseries 80} (2020) 627}
  [\href{https://arxiv.org/abs/1811.06387}{{\ttfamily 1811.06387}}].

\bibitem{Cai:2018dig}
R.-g. Cai, S.~Pi and M.~Sasaki, \emph{{Gravitational Waves Induced by
  non-Gaussian Scalar Perturbations}},
  \href{https://doi.org/10.1103/PhysRevLett.122.201101}{\emph{Phys. Rev. Lett.}
  {\bfseries 122} (2019) 201101}
  [\href{https://arxiv.org/abs/1810.11000}{{\ttfamily 1810.11000}}].

\bibitem{Jung:2019fcs}
S.~Jung and T.~Kim, \emph{{Gamma-ray burst lensing parallax: Closing the
  primordial black hole dark matter mass window}},
  \href{https://doi.org/10.1103/PhysRevResearch.2.013113}{\emph{Phys. Rev.
  Res.} {\bfseries 2} (2020) 013113}
  [\href{https://arxiv.org/abs/1908.00078}{{\ttfamily 1908.00078}}].

\bibitem{Bai:2018bej}
Y.~Bai and N.~Orlofsky, \emph{{Microlensing of X-ray Pulsars: a Method to
  Detect Primordial Black Hole Dark Matter}},
  \href{https://doi.org/10.1103/PhysRevD.99.123019}{\emph{Phys. Rev. D}
  {\bfseries 99} (2019) 123019}
  [\href{https://arxiv.org/abs/1812.01427}{{\ttfamily 1812.01427}}].

\bibitem{Wang:2019kzb}
Y.-F. Wang, Q.-G. Huang, T.~G. Li and S.~Liao, \emph{{Searching for primordial
  black holes with stochastic gravitational-wave background in the space-based
  detector frequency band}},
  \href{https://doi.org/10.1103/PhysRevD.101.063019}{\emph{Phys. Rev. D}
  {\bfseries 101} (2020) 063019}
  [\href{https://arxiv.org/abs/1910.07397}{{\ttfamily 1910.07397}}].

\bibitem{Dror:2019twh}
J.~A. Dror, H.~Ramani, T.~Trickle and K.~M. Zurek, \emph{{Pulsar Timing Probes
  of Primordial Black Holes and Subhalos}},
  \href{https://doi.org/10.1103/PhysRevD.100.023003}{\emph{Phys. Rev. D}
  {\bfseries 100} (2019) 023003}
  [\href{https://arxiv.org/abs/1901.04490}{{\ttfamily 1901.04490}}].

\bibitem{Guo:2017njn}
H.-K. Guo, J.~Shu and Y.~Zhao, \emph{{Using LISA-like Gravitational Wave
  Detectors to Search for Primordial Black Holes}},
  \href{https://doi.org/10.1103/PhysRevD.99.023001}{\emph{Phys. Rev. D}
  {\bfseries 99} (2019) 023001}
  [\href{https://arxiv.org/abs/1709.03500}{{\ttfamily 1709.03500}}].

\bibitem{Dutta:2020lqc}
B.~Dutta, A.~Kar and L.~E. Strigari, \emph{{Constraints on MeV dark matter and
  primordial black holes: Inverse Compton signals at the SKA}},
  \href{https://doi.org/10.1088/1475-7516/2021/03/011}{\emph{JCAP} {\bfseries
  03} (2021) 011} [\href{https://arxiv.org/abs/2010.05977}{{\ttfamily
  2010.05977}}].

\bibitem{Kusenko:2020pcg}
A.~Kusenko, M.~Sasaki, S.~Sugiyama, M.~Takada, V.~Takhistov and E.~Vitagliano,
  \emph{{Exploring Primordial Black Holes from the Multiverse with Optical
  Telescopes}},
  \href{https://doi.org/10.1103/PhysRevLett.125.181304}{\emph{Phys. Rev. Lett.}
  {\bfseries 125} (2020) 18}
  [\href{https://arxiv.org/abs/2001.09160}{{\ttfamily 2001.09160}}].

\bibitem{Sugiyama:2020roc}
S.~Sugiyama, V.~Takhistov, E.~Vitagliano, A.~Kusenko, M.~Sasaki and M.~Takada,
  \emph{{Testing Stochastic Gravitational Wave Signals from Primordial Black
  Holes with Optical Telescopes}},
  \href{https://doi.org/10.1016/j.physletb.2021.136097}{\emph{Phys. Lett. B}
  {\bfseries 814} (2021) 136097}
  [\href{https://arxiv.org/abs/2010.02189}{{\ttfamily 2010.02189}}].

\bibitem{Bhaumik:2020dor}
N.~Bhaumik and R.~K. Jain, \emph{{Stochastic induced gravitational waves and
  lowest mass limit of primordial black holes with the effects of reheating}},
  \href{https://arxiv.org/abs/2009.10424}{{\ttfamily 2009.10424}}.

\bibitem{Page:1976df}
D.~N. Page, \emph{{Particle Emission Rates from a Black Hole: Massless
  Particles from an Uncharged, Nonrotating Hole}},
  \href{https://doi.org/10.1103/PhysRevD.13.198}{\emph{Phys. Rev.} {\bfseries
  D13} (1976) 198}.

\bibitem{Page:1976ki}
D.~N. Page, \emph{{Particle Emission Rates from a Black Hole. 2. Massless
  Particles from a Rotating Hole}},
  \href{https://doi.org/10.1103/PhysRevD.14.3260}{\emph{Phys. Rev.} {\bfseries
  D14} (1976) 3260}.

\bibitem{MacGibbon:2007yq}
J.~H. MacGibbon, B.~J. Carr and D.~N. Page, \emph{{Do Evaporating Black Holes
  Form Photospheres?}},
  \href{https://doi.org/10.1103/PhysRevD.78.064043}{\emph{Phys. Rev.}
  {\bfseries D78} (2008) 064043}
  [\href{https://arxiv.org/abs/0709.2380}{{\ttfamily 0709.2380}}].

\bibitem{1980AA....81..263O}
P.~N. {Okele} and M.~J. {Rees}, \emph{{Observational consequences of positron
  production by evaporating black holes}}, {\emph{Astronomy and Astrophysics}
  {\bfseries 81} (1980) 263}.

\bibitem{okeke1980primary}
P.~Okeke, \emph{The primary source and the fates of galactic positrons},
  {\emph{Astrophysics and Space Science} {\bfseries 71} (1980) 371}.

\bibitem{1991ApJ...371..447M}
J.~H. {MacGibbon} and B.~J. {Carr}, \emph{{Cosmic Rays from Primordial Black
  Holes}}, \href{https://doi.org/10.1086/169909}{\emph{Astrophysical Journal}
  {\bfseries 371} (1991) 447}.

\bibitem{Bambi:2008kx}
C.~Bambi, A.~D. Dolgov and A.~A. Petrov, \emph{{Primordial black holes and the
  observed Galactic 511-keV line}},
  \href{https://doi.org/10.1016/j.physletb.2009.10.053,
  10.1016/j.physletb.2008.10.057}{\emph{Phys. Lett.} {\bfseries B670} (2008)
  174} [\href{https://arxiv.org/abs/0801.2786}{{\ttfamily 0801.2786}}].

\bibitem{Stocker:2018avm}
P.~St\"ocker, M.~Kr\"amer, J.~Lesgourgues and V.~Poulin, \emph{{Exotic energy
  injection with ExoCLASS: Application to the Higgs portal model and
  evaporating black holes}},
  \href{https://doi.org/10.1088/1475-7516/2018/03/018}{\emph{JCAP} {\bfseries
  03} (2018) 018} [\href{https://arxiv.org/abs/1801.01871}{{\ttfamily
  1801.01871}}].

\bibitem{Acharya:2020jbv}
S.~K. Acharya and R.~Khatri, \emph{{CMB and BBN constraints on evaporating
  primordial black holes revisited}},
  \href{https://doi.org/10.1088/1475-7516/2020/06/018}{\emph{JCAP} {\bfseries
  06} (2020) 018} [\href{https://arxiv.org/abs/2002.00898}{{\ttfamily
  2002.00898}}].

\bibitem{Keith:2020jww}
C.~Keith, D.~Hooper, N.~Blinov and S.~D. McDermott, \emph{{Constraints on
  Primordial Black Holes From Big Bang Nucleosynthesis Revisited}},
  \href{https://doi.org/10.1103/PhysRevD.102.103512}{\emph{Phys. Rev. D}
  {\bfseries 102} (2020) 103512}
  [\href{https://arxiv.org/abs/2006.03608}{{\ttfamily 2006.03608}}].

\bibitem{Kouvaris:2018wnh}
C.~Kouvaris, P.~Tinyakov and M.~H. Tytgat, \emph{{NonPrimordial Solar Mass
  Black Holes}},
  \href{https://doi.org/10.1103/PhysRevLett.121.221102}{\emph{Phys. Rev. Lett.}
  {\bfseries 121} (2018) 221102}
  [\href{https://arxiv.org/abs/1804.06740}{{\ttfamily 1804.06740}}].

\bibitem{Dasgupta:2020mqg}
B.~Dasgupta, R.~Laha and A.~Ray, \emph{{Low Mass Black Holes from Dark Core
  Collapse}}, \href{https://doi.org/10.1103/PhysRevLett.126.141105}{\emph{Phys.
  Rev. Lett.} {\bfseries 126} (2021) 141105}
  [\href{https://arxiv.org/abs/2009.01825}{{\ttfamily 2009.01825}}].

\bibitem{Shandera:2018xkn}
S.~Shandera, D.~Jeong and H.~S.~G. Gebhardt, \emph{{Gravitational Waves from
  Binary Mergers of Subsolar Mass Dark Black Holes}},
  \href{https://doi.org/10.1103/PhysRevLett.120.241102}{\emph{Phys. Rev. Lett.}
  {\bfseries 120} (2018) 241102}
  [\href{https://arxiv.org/abs/1802.08206}{{\ttfamily 1802.08206}}].

\bibitem{2019BAAS...51g.245M}
J.~{McEnery}, A.~{van der Horst}, A.~{Dominguez}, A.~{Moiseev}, A.~{Marcowith},
  A.~{Harding} et~al., \emph{{All-sky Medium Energy Gamma-ray Observatory:
  Exploring the Extreme Multimessenger Universe}},  in \emph{Bulletin of the
  American Astronomical Society}, vol.~51, p.~245, Sept., 2019,
  \href{https://arxiv.org/abs/1907.07558}{{\ttfamily 1907.07558}}.

\bibitem{MacGibbon:1990zk}
J.~H. MacGibbon and B.~R. Webber, \emph{{Quark and gluon jet emission from
  primordial black holes: The instantaneous spectra}},
  \href{https://doi.org/10.1103/PhysRevD.41.3052}{\emph{Phys. Rev.} {\bfseries
  D41} (1990) 3052}.

\bibitem{MacGibbon:1991tj}
J.~H. MacGibbon, \emph{{Quark and gluon jet emission from primordial black
  holes. 2. The Lifetime emission}},
  \href{https://doi.org/10.1103/PhysRevD.44.376}{\emph{Phys. Rev.} {\bfseries
  D44} (1991) 376}.

\bibitem{Arbey:2019mbc}
A.~Arbey and J.~Auffinger, \emph{{BlackHawk: A public code for calculating the
  Hawking evaporation spectra of any black hole distribution}},
  \href{https://doi.org/10.1140/epjc/s10052-019-7161-1}{\emph{Eur. Phys. J. C}
  {\bfseries 79} (2019) 693}
  [\href{https://arxiv.org/abs/1905.04268}{{\ttfamily 1905.04268}}].

\bibitem{Bartels:2017dpb}
R.~Bartels, D.~Gaggero and C.~Weniger, \emph{{Prospects for indirect dark
  matter searches with MeV photons}},
  \href{https://doi.org/10.1088/1475-7516/2017/05/001}{\emph{JCAP} {\bfseries
  1705} (2017) 001} [\href{https://arxiv.org/abs/1703.02546}{{\ttfamily
  1703.02546}}].

\bibitem{Carr:2009jm}
B.~Carr, K.~Kohri, Y.~Sendouda and J.~Yokoyama, \emph{{New cosmological
  constraints on primordial black holes}},
  \href{https://doi.org/10.1103/PhysRevD.81.104019}{\emph{Phys. Rev. D}
  {\bfseries 81} (2010) 104019}
  [\href{https://arxiv.org/abs/0912.5297}{{\ttfamily 0912.5297}}].

\bibitem{Niikura:2017zjd}
H.~Niikura et~al., \emph{{Microlensing constraints on primordial black holes
  with Subaru/HSC Andromeda observations}},
  \href{https://doi.org/10.1038/s41550-019-0723-1}{\emph{Nat. Astron.}
  {\bfseries 3} (2019) 524} [\href{https://arxiv.org/abs/1701.02151}{{\ttfamily
  1701.02151}}].

\bibitem{1994A&A...292...82S}
A.~W. {Strong}, K.~{Bennett}, H.~{Bloemen}, R.~{Diehl}, W.~{Hermsen},
  D.~{Morris} et~al., \emph{{Diffuse continuum gamma rays from the Galaxy
  observed by COMPTEL.}}, {\emph{Astronomy and Astrophysics} {\bfseries 292}
  (1994) 82}.

\bibitem{Strong:1998ck}
A.~Strong, H.~Bloemen, R.~Diehl, W.~Hermsen and V.~Schoenfelder, \emph{{Comptel
  skymapping: A New approach using parallel computing}}, {\emph{Astrophys.
  Lett. Commun.} {\bfseries 39} (1999) 209}
  [\href{https://arxiv.org/abs/astro-ph/9811211}{{\ttfamily
  astro-ph/9811211}}].

\bibitem{Strong:2011pa}
{\scshape Fermi-LAT} collaboration, A.~Strong, \emph{{Interstellar gamma rays
  and cosmic rays: new insights from Fermi-LAT AND INTEGRAL}},  in
  \emph{{ICATPP Conference on Cosmic Rays for Particle and Astroparticle
  Physics}}, pp.~473--481, 1, 2011,
  \href{https://arxiv.org/abs/1101.1381}{{\ttfamily 1101.1381}},
  \href{https://doi.org/10.1142/9789814329033_0059}{DOI}.

\bibitem{1997AIPC..410.1223W}
K.~{Watanabe}, D.~H. {Hartmann}, M.~D. {Leising}, L.~S. {The}, G.~H. {Share}
  and R.~L. {Kinzer}, \emph{{The Cosmic {\ensuremath{\gamma}}-ray Background
  from supernovae}},  in \emph{Proceedings of the Fourth Compton Symposium}
  (C.~D. {Dermer}, M.~S. {Strickman} and J.~D. {Kurfess}, eds.), vol.~410 of
  \emph{American Institute of Physics Conference Series}, pp.~1223--1227, May,
  1997, \href{https://doi.org/10.1063/1.53933}{DOI}.

\bibitem{1975Ap&SS..32L...1F}
Y.~{Fukada}, S.~{Hayakawa}, M.~{Ikeda}, I.~{Kasahara}, F.~{Makino} and
  Y.~{Tanaka}, \emph{{Rocket Observation of Energy Spectrum of Diffuse Hard
  X-Rays}}, \href{https://doi.org/10.1007/BF00646232}{\emph{Astrophysics and
  Space Science} {\bfseries 32} (1975) L1}.

\bibitem{Gruber:1999yr}
D.~Gruber, J.~Matteson, L.~Peterson and G.~Jung, \emph{{The spectrum of diffuse
  cosmic hard x-rays measured with heao-1}},
  \href{https://doi.org/10.1086/307450}{\emph{Astrophys. J.} {\bfseries 520}
  (1999) 124} [\href{https://arxiv.org/abs/astro-ph/9903492}{{\ttfamily
  astro-ph/9903492}}].

\bibitem{1997ApJ...475..361K}
R.~L. {Kinzer}, G.~V. {Jung}, D.~E. {Gruber}, J.~L. {Matteson}, {Peterson} and
  {L.~E.}, \emph{{Diffuse Cosmic Gamma Radiation Measured by HEAO 1}},
  \href{https://doi.org/10.1086/303507}{\emph{Astrophysical Journal} {\bfseries
  475} (1997) 361}.

\bibitem{Weidenspointner:2000aq}
G.~Weidenspointner et~al., \emph{{The comptel instrumental line background}},
  \href{https://doi.org/10.1063/1.1303269}{\emph{AIP Conf. Proc.} {\bfseries
  510} (2000) 581} [\href{https://arxiv.org/abs/astro-ph/0012332}{{\ttfamily
  astro-ph/0012332}}].

\bibitem{Edwards:2017mnf}
T.~D.~P. Edwards and C.~Weniger, \emph{{A Fresh Approach to Forecasting in
  Astroparticle Physics and Dark Matter Searches}},
  \href{https://doi.org/10.1088/1475-7516/2018/02/021}{\emph{JCAP} {\bfseries
  02} (2018) 021} [\href{https://arxiv.org/abs/1704.05458}{{\ttfamily
  1704.05458}}].

\bibitem{Edwards:2017kqw}
T.~D.~P. Edwards and C.~Weniger, \emph{{swordfish: Efficient Forecasting of New
  Physics Searches without Monte Carlo}},
  \href{https://arxiv.org/abs/1712.05401}{{\ttfamily 1712.05401}}.

\bibitem{doornhein2018uses}
M.~T. Doornhein, \emph{{Uses and Limitations of Fisher Forecasting in Setting
  Upper Limits on the Interaction Strength of Dark Matter, MSc thesis}},  2018.

\bibitem{Navarro:1996gj}
J.~F. Navarro, C.~S. Frenk and S.~D.~M. White, \emph{{A Universal density
  profile from hierarchical clustering}},
  \href{https://doi.org/10.1086/304888}{\emph{Astrophys. J.} {\bfseries 490}
  (1997) 493} [\href{https://arxiv.org/abs/astro-ph/9611107}{{\ttfamily
  astro-ph/9611107}}].

\bibitem{Ng:2013xha}
K.~C.~Y. Ng, R.~Laha, S.~Campbell, S.~Horiuchi, B.~Dasgupta, K.~Murase et~al.,
  \emph{{Resolving small-scale dark matter structures using multisource
  indirect detection}},
  \href{https://doi.org/10.1103/PhysRevD.89.083001}{\emph{Phys. Rev. D}
  {\bfseries 89} (2014) 083001}
  [\href{https://arxiv.org/abs/1310.1915}{{\ttfamily 1310.1915}}].

\bibitem{Winkler:2003nn}
C.~Winkler et~al., \emph{{The INTEGRAL mission}},
  \href{https://doi.org/10.1051/0004-6361:20031288}{\emph{Astron. Astrophys.}
  {\bfseries 411} (2003) L1}.

\bibitem{Atwood:2009ez}
{\scshape Fermi-LAT} collaboration, W.~B. Atwood et~al., \emph{{The Large Area
  Telescope on the Fermi Gamma-ray Space Telescope Mission}},
  \href{https://doi.org/10.1088/0004-637X/697/2/1071}{\emph{Astrophys. J.}
  {\bfseries 697} (2009) 1071}
  [\href{https://arxiv.org/abs/0902.1089}{{\ttfamily 0902.1089}}].

\bibitem{1993ApJS...86..657S}
V.~{Schoenfelder}, H.~{Aarts}, K.~{Bennett}, H.~{de Boer}, J.~{Clear},
  W.~{Collmar} et~al., \emph{{Instrument Description and Performance of the
  Imaging Gamma-Ray Telescope COMPTEL aboard the Compton Gamma-Ray
  Observatory}}, \href{https://doi.org/10.1086/191794}{\emph{Astrophysical
  Journal, Supplement} {\bfseries 86} (1993) 657}.

\bibitem{Coe:2009xf}
D.~Coe, \emph{{Fisher Matrices and Confidence Ellipses: A Quick-Start Guide and
  Software}},  \href{https://arxiv.org/abs/0906.4123}{{\ttfamily 0906.4123}}.

\end{thebibliography}\endgroup
 
  \end{document}